\documentclass[12pt,preprint]{aastex}

\newcommand{\htwo}{H\,{\sc ii}}
\usepackage{epsfig}
\usepackage{color}
\begin{document}

\title{A Classification Scheme for Young Stellar Objects Using the
  {\it WIDE-FIELD INFRARED SURVEY EXPLORER} {\it AllWISE} Catalog:
  Revealing Low-Density Star Formation in the Outer Galaxy}

\author{X.~P. Koenig,\altaffilmark{1} D.~T. Leisawitz\altaffilmark{2}}
\altaffiltext{1}{Yale University}
\altaffiltext{2}{NASA Goddard Space Flight Center, Greenbelt, MD 20771, USA}

\begin{abstract}
We present an assessment of the performance of {\it WISE} and the {\it
  AllWISE} data release in a section of the Galactic Plane. We lay out
an approach to increasing the reliability of point source photometry
extracted from the {\it AllWISE} catalog in Galactic Plane regions
using parameters provided in the catalog. We use the resulting catalog
to construct a new, revised young star detection and classification
scheme combining {\it WISE} and {\it 2MASS} near and mid-infrared
colors and magnitudes and test it in a section of the Outer Milky
Way. The clustering properties of the candidate Class I and II stars
using a nearest neighbor density calculation and the two-point
correlation function suggest that the majority of stars do form in
massive star forming regions, and any isolated mode of star formation
is at most a small fraction of the total star forming output of the
Galaxy. We also show that the isolated component may be very small and
could represent the tail end of a single mechanism of star formation
in line with models of molecular cloud collapse with supersonic
turbulence and not a separate mode all to itself.
\end{abstract}


\section{Introduction}
The Wide-field Infrared Survey Explorer ({\it WISE}) mapped the whole
sky in 4 mid-infrared bands at 3.4, 4.6, 12, and 22~$\micron$
\citep{wright10}. The location of these bands in the mid-infrared
matches where the excess emission from cooler circumstellar
disk/envelope material in young stars begins to become significant in
relation to the stellar photosphere. This fact means that {\it WISE}
can be readily used as a tool to find and classify young stellar
objects (YSOs), in a similar way to work carried out with {\it
  Spitzer} (Allen et~al.\ 2004, Gutermuth et~al.\ 2008, 2009 and
others). However, given its specific set of science goals, the {\it
  WISE} pipeline source extraction process was not optimized for the
areas where these objects are most commonly found, namely the Galactic
Plane. There, YSOs are typically found in regions that are bright in
the mid-infrared from thermal dust emission or excited line emission
from PAH dust, or regions that are dark as in the case of infrared
dark clouds \citep{rathb06} or in dense clustered regions. These facts
necessitate an assessment of {\it WISE}'s performance in such regions
before we set out schemes to achieve our goal of finding YSOs. In a
recent paper, \citet{koenig12} (hereafter: K12) described a {\it
  WISE-2MASS} \citep[Two Micron All Sky Survey, see][]{skrut06} YSO
identification and classification scheme and applied it to photometry
they extracted from their own point source catalogs extracted from
custom-made {\it WISE} mosaic images. While this was practical for a
`pointed' survey such as theirs, a general scheme that can be applied
to arbitrarily located and sized portions of the sky is necessary when
computational limitations make the construction of huge mosaics
prohibitive. An examination of the reliability of the {\it WISE}
catalog in the Galactic Plane should also be generally useful to the
science community and is not yet covered in great detail in existing
online documentation for the survey.

The properties of the public {\it WISE} source catalogs stem in part
from the choice of source detection run on the four-band imaging data
from the telescope. The choice of source detection itself was
motivated by the needs of the driving science goals of the mission: to
study infrared-bright galaxies, to find brown dwarfs and to study
near-Earth asteroids. The online explanatory
supplement\footnote{http://wise2.ipac.caltech.edu/docs/release/allwise/expsup/}
and \citet{marsh12} describe the {\it WISE} source detection method in
detail. The process is based on the thresholding of a `detection'
image derived from a set of matched filter images in the relevant
bands. The algorithm calculates an optimal matched filter at each
wavelength and combines the resulting single-band images in quadrature
to produce a detection image in units of the local standard deviation
of noise, and then searches this image for local maxima. It then
further deblends candidate point sources by profile-fitting on each
individual frame for all bands simultaneously. This process avoids the
need for bandmerging and allows for improved detection of sources in
the longer wavelength {\it WISE} bands. However in Galactic
star-forming regions it also has the negative consequence of
generating detections of the nebular background with apparently good
photometric parameters (in particular: signal to noise and photometric
uncertainty) in some bands in accompaniment to genuine point source
detections in the other bands. In this paper we will document the
performance of the {\it AllWISE} catalog in the Galactic Plane and
outline approaches to mitigating fake source contamination. We will
apply one such approach in order to present a refined version of the
YSO classification scheme of K12. This paper will also account in more
detail for the various astrophysical sources of contamination that can
be misclassified as YSOs by schemes based on infrared colors, for
example, Asymptotic Giant Branch (AGB) and Classical Be stars
\citep[CBe: rapidly rotating B-type main sequence stars with a
  geometrically-flattened decretion disk,][]{rivin13}. It should be
noted that the {\it AllWISE} release differs from prior {\it WISE}
releases in how it handles background subtraction and other details of
photometric extraction, as described in the explanatory
supplement. While the overall approach we lay out in this paper for
fake source mitigation and young star identification would be
applicable to the Preliminary and All-Sky releases, the precise
details would thus differ slightly.

\section{Data}
In this paper we make use of the {\it AllWISE} processing of the {\it
  WISE} survey data. As part of the online point source catalog, a
match to the {\it 2MASS} point-source catalog is automatically
included and we include these data as well.

\subsection{Properties of {\it WISE} and the {\it AllWISE} catalog in the Galactic Plane}
In order to understand how {\it WISE} in general and specifically the
{\it AllWISE} point source extraction perform in Galactic star forming
regions, we analyzed two test regions covering parts of the W3 and W5
giant molecular clouds and part of the W4 region
\citep{westerhout58}. Both the W3/W4 and the W5 test fields encompass
on-cloud, bright, nebular emission and off-cloud, lower background
regions in their neighboring, ionized \htwo\ regions. The W3/W4 field
(from here on referred to as just the `W3' field) additionally
incorporates a dark lane or infrared dark cloud. To clarify the
following discussion, we adopt lower case italics to refer to the four
{\it WISE} bands, $w1$, $w2$, $w3$ and $w4$ and upper case W3, W4 and
W5 to refer to the massive star forming regions.

To estimate the maximum potential completeness of {\it WISE} in
Galactic star forming regions like W3, W4 and W5, we start by
constructing a `truth' catalog for each band for each field from the
large {\it WISE} mosaics of W3, 4 and 5 made by K12 from raw {\it
  WISE} frames, since these have slightly higher resolution than the
{\it WISE} atlas tiles available online. This property of the {\it
  WISE} image atlas tiles is due to the co-addition process that
slightly smoothes these images to optimize them for point source
detection, but degrades their angular resolution by a factor of
$\sqrt{2}$ (see the {\it AllWISE} Explanatory Supplement, section
IV.4.l). We search our images for point sources in the test fields by
eye, verifying these with reference to the {\it Spitzer} images of
\citet{koenig08} for W5 and mosaics from the {\it Spitzer} Heritage
Archive\footnote{http://sha.ipac.caltech.edu/applications/Spitzer/SHA/}
for W3. The {\it Spitzer} IRAC 3.6 and 4.5~$\micron$ and MIPS
24~$\micron$ images are a close match to {\it WISE} bands 1, 2 and
4. The IRAC 8~$\micron$ band is not an exact match to {\it WISE} band
3 at 12~$\micron$, but is close enough in wavelength that point
sources bright in these {\it Spitzer} images are similarly evident in
{\it WISE} 12~$\micron$ images, allowing a determination if an
apparent {\it WISE} band 3 point source is in fact a nebular emission
knot when seen at the higher sensitivity and resolution of {\it
  Spitzer}. In the following discussion we refer to objects in this
list and their matched entries in the {\it WISE} photometric database
as `real' sources, while objects in the database with no match in the
truth catalog in a given band are referred to as `fake' or `spurious'
sources.

Since the point source density varies strongly with waveband, the size
of the region surveyed by eye is different between
bands. Table~\ref{tab:fields} lists the test field locations and
source sample sizes.

\begin{deluxetable}{cccccc}
\tablewidth{0pt} 
\tabletypesize{\scriptsize}
\tablecaption{$WISE$ Test Fields and Sample Sizes}
\tablehead{
\colhead{Cloud} & \colhead{{\it WISE} band} & \colhead{Center} & \colhead{Dimensions} & \colhead{$N_{SRC}$} & \colhead{$N_{non-null}$}}
\startdata
W~3 & 1 & 2:28:48 +61:32:00.0 & 17.2$\times$10.6 & 1713 & 501 \\
 & 2 & 2:28:48 +61:32:00.0 & 17.2$\times$10.6 & 1646 & 488 \\
 & 3 & 2:27:30 +61:25:45.0 & 46$\times$45 & 704 & 464 \\
 & 4 & 2:26:00 +61:12:20.0 & 87$\times$96 & 401 & 333 \\
W~5 & 1 & 2:55:10 +60:40:15.0 & 14.9$\times$10.65 & 1633 & 603 \\
 & 2 & 2:55:10 +60:40:15.0 & 14.9$\times$10.65 & 1309 & 583 \\
 & 3 & 2:55:00 +60:40:15.0 & 44.5$\times$44.5 & 710 & 549 \\
 & 4 & 2:52:50 +60:39:35.0 & 90$\times$90 & 399 & 335
\label{tab:fields}
\enddata
\tablecomments{Coordinate centers are J2000.0, field dimensions are in
  arcminutes. $N_{SRC}$ gives the number of point sources visible in
  each field. $N_{non-null}$ is the subset of $N_{SRC}$ that are found
  within 2$\arcsec$ (for $w1$ and $w2$) or 3$\arcsec$ (for $w3$ and
  $w4$) in the {\it AllWISE} catalog with non-null photometric error.}
\end{deluxetable}

\subsection{Completeness of {\it WISE} and {\it AllWISE}}
The nominal 95\% completeness listed for the {\it WISE} survey in the
Explanatory Supplement is appropriate for high Galactic latitude
fields with $|b|>30\degr$. In magnitudes, the values are: $w1$=17.1,
$w2$=15.7, $w3$=11.6 and $w4$=7.7. In the Galactic Plane, as already
mentioned, star forming regions contain bright, structured emission
from PAH molecules and dust grains excited and heated by radiation
from nearby and embedded young stars \citep{li01} and extinction
features like infrared dark clouds. The point source density is also
higher in the Galactic Plane and stars are frequently found in dense
clusters. As described in Appendix A.1, we determine a measure of the
magnitude at which 90\% of sources that were detected by {\it Spitzer}
in our test fields are found in the {\it WISE}
images. Table~\ref{tab:potencomp} summarizes the results of this
calculation. In regions of bright background, the intrinsic detection
limit of {\it WISE} is lower, thus we generate a completeness estimate
for regions of high and low sky background in each of our test
regions. Because the {\it WISE} source list was determined by eye with
no deblending capability, the true completeness may be slightly better
than the quoted values.

\begin{deluxetable}{ccccc}
\tablewidth{0pt} 
\tabletypesize{\scriptsize}
\tablecaption{$WISE$ Potential 90\% Completeness Limits}
\tablehead{\colhead{} & \multicolumn{2}{c}{W3} & \multicolumn{2}{c}{W5} \\
\colhead{Band} & \colhead{Low Sky} & \colhead{High Sky} & \colhead{Low Sky} & \colhead{High Sky}}
\startdata
1 & 14.75 & 12.25 & 14.75 & 14.75  \\
2 & 14.75 & 11.75 & 14.25 & 14.25  \\
3 & 10.75 & 8.25 & 11.25 & 9.25 \\
4 & 6.25 & 3.25 & 7.25 & 5.75 
\label{tab:potencomp}
\enddata
\end{deluxetable}

The {\it AllWISE} source extraction pipeline is only able to retrieve
a subset of the sources visible in Galactic Plane {\it WISE} images,
because of the issues described earlier. Table~\ref{tab:fields} lists
the number of truth catalog sources found in the {\it AllWISE} catalog
within 2$\arcsec$ (for $w1$ and $w2$) or 3$\arcsec$ (for $w3$ and
$w4$) with non-null photometric error (the $w?sigmpro$ column in the
online table). We calculate the 90\% magnitude completeness limit of
the {\it AllWISE} source extraction in these test fields by comparing
histograms of source magnitude of {\it AllWISE} objects that match the
truth catalog (so as to filter off spurious detections), to the {\it
  Spitzer} catalogs for the same regions. The completeness estimates
for {\it AllWISE} are presented in Table~\ref{tab:allwcomp}.

\begin{deluxetable}{ccccc}
\tablewidth{0pt} 
\tabletypesize{\scriptsize}
\tablecaption{$AllWISE$ 90\% Completeness Limits}
\tablehead{\colhead{} & \multicolumn{2}{c}{W3} & \multicolumn{2}{c}{W5} \\
\colhead{Band} & \colhead{Low Sky} & \colhead{High Sky} & \colhead{Low Sky} & \colhead{High Sky}}
\startdata
1 & 13.75 & 11.75 & 13.25 & 13.25  \\
2 & 13.75 & 10.75 & 13.25 & 13.25  \\
3 & 9.25 & 8.25 & 10.25 & 9.25 \\
4 & 5.25 & 3.75 & 7.25 & 5.75 
\label{tab:allwcomp}
\enddata
\end{deluxetable}

\subsection{Fake Source Contamination in {\it AllWISE}}
The `truth' catalog allows us to estimate the rates of contamination
by fake sources in the {\it AllWISE} point source catalog by band, on
and off cloud in areas of high and low sky background respectively. In
each test field we search for objects within 2$\arcsec$ (for $w1$ and
$w2$) or 3$\arcsec$ (for $w3$ and $w4$) of our by-eye source lists. In
Table~\ref{tab:contam} we present the number of objects that match the
truth catalog with signal to noise $\geq$5 in that band, in high and
low sky regions, versus the number of objects with signal to noise
$\geq$5 in {\it AllWISE} that have no visible point source in that
band in the test fields. In low sky regions, the contamination in $w1$
and $w2$ is $\approx$10\%, but in high sky regions about 1/3 of
sources above a signal to noise ratio (SNR) of 5 appear spurious. In
the $w3$ and $w4$ bands, the vast majority of sources above SNR=5 in
the {\it AllWISE} catalog are likely spurious detections and simply
represent upper limit measurements of the nebular background. At the
same time, adding up the total number of real sources in
Table~\ref{tab:contam}, or looking at the final column in
Table~\ref{tab:fields} also shows that in bands 1 and 2 only about 1/3
of all the sources visible in the images are detected by {\it AllWISE}
at all (compare $N_{SRC}$ and $N_{non-null}$ in
Table~\ref{tab:fields}). In band 3, {\it AllWISE} picks up 50--60\% of
truth catalog sources, while in band 4, {\it AllWISE} finds about 3/4
of truth catalog sources. We now look for parameters in the {\it
  AllWISE} data release that can help mitigate its unreliability as a
YSO finding tool if used incautiously. At the same time we attempt to
maintain a high retrieval rate of real sources.

We note that photometric quality flags that designate sources as upper
limit detections exist in the {\it AllWISE} catalog in the $ph\_qual$
column. But by requiring sources have non-null photometric error as a
first, basic requirement, we already implicitly allow $ph\_qual$ to be
only `A', `B' or `C' in that band, or equivalently SNR$>$2. In the key
challenge of suppressing the huge amount of fake detections in bands 3
and 4, even requiring only $ph\_qual$ `A' sources only reduces the
contamination rate to ~85\% and 94\% respectively. In bands 1 and 2,
the contamination rate is already quite low (Table~\ref{tab:contam})
and would only drop by a few percent by requiring $ph\_qual$ = `A'
only.

\begin{deluxetable}{ccccccccc}
\tablewidth{0pt} 
\tabletypesize{\scriptsize}
\tablecaption{{\it AllWISE} Fake Source Contamination}
\tablehead{\colhead{} & \multicolumn{4}{c}{High Sky Bkgd. Contamination} & \multicolumn{4}{c}{Low Sky Bkgd. Contamination} \\
\colhead{Band} & \multicolumn{2}{c}{W3} & \multicolumn{2}{c}{W5} & \multicolumn{2}{c}{W3} & \multicolumn{2}{c}{W5}\\
\colhead{ } & \colhead{$N_{real}/N_{fake}$} & \colhead{$\% fake$} & \colhead{$N_{real}/N_{fake}$} & \colhead{$\% fake$} & \colhead{$N_{real}/N_{fake}$} & \colhead{$\% fake$} & \colhead{$N_{real}/N_{fake}$} & \colhead{$\% fake$}  }
\startdata
1 & 400/209 & 34.3 & 143/79 & 35.6 & 94/19 & 16.8 & 460/62 & 11.9 \\
2 & 344/176 & 33.8 & 163/84 & 34.0 & 122/15 & 10.9 & 414/32 & 7.2  \\
3 & 208/2457 & 92.2 & 158/2081 & 92.9 & 168/485 & 74.3 & 293/62 & 75.8 \\
4 & 135/5526 & 97.6 & 169/5554 & 97.0 & 168/2435 & 93.5 & 133/2379 & 94.7 
\label{tab:contam}
\enddata
\end{deluxetable}

\section{{\it AllWISE} Fake Source Mitigation} \label{sec:overallmit}

\subsection{{\it AllWISE} Catalog Parameters and their Properties in Real and Fake Sources}

\subsubsection{Signal-to-noise and Chi-squared} \label{sec:snrchi}
The signal to noise ($w?snr$) and reduced chi-squared ($w?rchi2$)
parameters given in the {\it AllWISE} photometric catalog give a
strong discriminating power between real and fake point sources. The
behavior of $w?rchi2$ versus $w?snr$ in the two test fields in the
four {\it WISE} bands is shown in Figure~\ref{fig:SN-Chi} and in the
Appendix: Figure~\ref{fig:Appendix-SN-Chi}. Hereafter we use
$\chi_{\nu}^2$ and SNR to refer to reduced chi-squared and signal to
noise respectively.

In general, the behavior seen in Fig.~\ref{fig:SN-Chi} is consistent
with the expectation that a true point source will have $\chi_{\nu}^2
\approx 1$ from the profile fitting procedure. Fake sources have
$\chi_{\nu}^2 \approx 1$ at low signal to noise, but start to deviate
away as the measured signal to noise increases. When examined in the
{\it WISE} images, real sources with $\chi_{\nu}^2 > 2$ are crowded by
nearby neighboring sources, are in regions of complex and/or bright
background emission, or are extended.

We apply the following cuts in SNR vs. $\chi_{\nu}^2$ space as shown
in Fig.~\ref{fig:SN-Chi}:

\begin{eqnarray}
  w1rchi2 & < & (w1snr - 3)/7 \\
  w2rchi2 & < & 0.1 \times w2snr - 0.3 \\
  w3rchi2 & < & 0.125 \times w3snr - 1 \\
  w4rchi2 & < & 0.2 \times w4snr - 2
\end{eqnarray}

to suppress the fake component. These criteria also allow some
fraction of the real sources with high $\chi_{\nu}^2$ to be
included. We tabulate the retrieval rate of real sources and the
residual contamination rate by spurious catalog entries in
Table~\ref{tab:SN-Chi}. In this and the following sub-sections our
starting point for the retrieval rate is the final column of
Table~\ref{tab:fields}, so we implicitly always require that {\it
  AllWISE} sources in a given band have non-null photometric
uncertainty $w?sigmpro$. For example, in band 1 in the W3 test field,
Table~\ref{tab:fields} shows that we find 501 real sources with
non-null $w?sigmpro$. Table~\ref{tab:SN-Chi} shows that $\approx$57\%
of these sources are retrieved by the band 1 SNR vs. $\chi_{\nu}^2$
cut, and 3\% of all the {\it AllWISE} sources that have non-null
$w?sigmpro$ and pass this cut are fake.  With these parameters it is
possible to suppress the contamination rate in any band down to
$<7$\%, although this success comes at a cost of cutting about 2/3 of
the real sources in bands 3 and 4.

\begin{deluxetable}{ccccc}
\tablewidth{0pt} 
\tabletypesize{\scriptsize}
\tablecaption{Signal to Noise and Chi-squared}
\tablehead{\colhead{} & \multicolumn{2}{c}{Real Retrieval Rate (\%)} & \multicolumn{2}{c}{Residual Fake (\% of total)} \\
\colhead{Band} & \colhead{W3} & \colhead{W5} & \colhead{W3} & \colhead{W5} }
\startdata
1 & 56.9 & 74.8 & 3.1 & 3.8 \\
2 & 60.0 & 78.3 & 5.5 & 4.8 \\
3 & 23.7 & 32.1 & 4.3 & 6.4 \\
4 & 41.1 & 34.3 & 1.4 & 6.5 
\label{tab:SN-Chi}
\enddata 
\tablecomments{Retrieval rates are given as the percentage of the real
  `truth' catalog sources found in {\it AllWISE} with non-null
  photometric error, that are retrieved by the $w?rchi2$ versus
  $w?snr$ selection criteria in $\S$~3.1.1. The residual fake columns
  are the percentage of all the sources allowed through by these
  criteria that are still fake/spurious.}
\end{deluxetable}

\begin{figure}
\centering
\includegraphics[width=12cm]{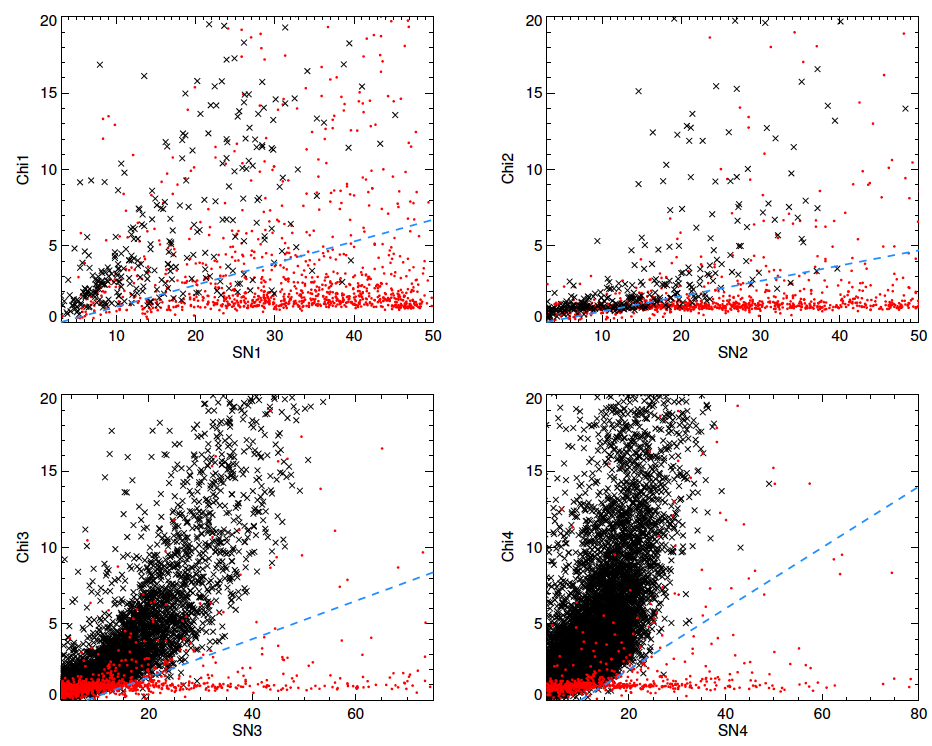}
\caption{The distribution in signal to noise and profile fit reduced
  chi-squared of truth catalog point sources found in {\it AllWISE}
  (red points) and unmatched, spurious {\it AllWISE} catalog entries
  (black $\times$-points) in the four {\it WISE}
  bands.\label{fig:SN-Chi} Blue dashed line shows the selection cuts
  we use in this paper.}
\end{figure}

\subsubsection{$w?nm$ and $w?m$}
The catalog parameters $w?nm$ (the number of profile-fit flux
measurements for a source with $w?snr > 3$) and $w?m$ (the number of
profile-fit flux measurements for a source) also have some potential
for separating real and fake sources in the catalog. We might expect
that a real point source would be detected with $w?snr > 3$ in a
higher fraction of the {\it WISE} frames. Figure~\ref{fig:wnm-wm} in
the Appendix shows the behavior of real and fake sources in these
parameters. Table~\ref{tab:wnm-wm} presents a summary of the results
of requiring $w?nm$/$w?m>0.2$. In practice, {\it WISE} band 4 shows
the largest percentage of fake sources with low $w4nm/w4m$ that might
be filtered off by such a search criterion.

\begin{deluxetable}{ccccc}
\tablewidth{0pt} 
\tabletypesize{\scriptsize}
\tablecaption{$w?nm$ and $w?m$}
\tablehead{\colhead{} & \multicolumn{2}{c}{Real Retrieval Rate} & \multicolumn{2}{c}{Residual Fake (\% of total)} \\
\colhead{Band} & \colhead{W3} & \colhead{W5} & \colhead{W3} & \colhead{W5} }
\startdata
1 & 99.8 & 100 & 32.7 & 19.1 \\
2 & 94.5 & 96.1 & 28.9 & 16.5 \\
3 & 87.5 & 83.8 & 89.1 & 87.6 \\
4 & 90.1 & 87.8 & 96.3 & 96.4 
\label{tab:wnm-wm}
\enddata 
\tablecomments{Retrieval rates are given as the percentage of real
  `truth' catalog sources found in {\it AllWISE} with non-null
  photometric error, that are retrieved with $w?nm$/$w?m>0.2$ in a
  given band and test field (see $\S$~3.1.2). The residual fake
  columns are the percentage of all the sources that are selected by
  these criteria that are still fake/spurious.}
\end{deluxetable}

\subsubsection{$nb$ and $na$}
The catalog flags $na$ and $nb$ indicate whether a source has been
actively deblended ($na=0$ means no, `1' indicates yes) and if so, how
many blend components were used ($nb$). Requiring $nb=1$ automatically
produces $na=0$. Setting $na=0$ does not produce $nb=1$ however, since
$nb$ can be greater than 1 if a source was passively
deblended. Table~\ref{tab:nb-na} summarizes the behavior of real and
fake sources by band and field. We list the fractions of sources with
$nb=1$ and non-null photometry in each band, noting however, that $nb$
is defined once for each catalog entry and is not separately computed
for each band. In general, real sources are more likely to be
unblended, but the level of suppression of fake sources in bands 3 and
4 is modest and in bands 1 and 2 comes at the cost of cutting about
30\% of real sources.

\begin{deluxetable}{ccccc}
\tablewidth{0pt} 
\tabletypesize{\scriptsize}
\tablecaption{$nb$ and $na$}
\tablehead{\colhead{} & \multicolumn{2}{c}{Real Retrieval Rate} & \multicolumn{2}{c}{Residual Fake (\% of total)} \\
\colhead{Band} & \colhead{W3} & \colhead{W5} & \colhead{W3} & \colhead{W5} }
\startdata
1 & 69.7 & 68.3 & 20.3 & 11.2 \\
2 & 68.9 & 68.8 & 18.0 & 8.4 \\
3 & 84.5 & 83.1 & 84.2 & 81.7 \\
4 & 84.1 & 78.5 & 96.3 & 96.3 
\label{tab:nb-na}
\enddata
\tablecomments{Retrieval rates are given as the percentage of real
  `truth' catalog sources found in {\it AllWISE} with non-null
  photometric error, that have $nb$=1 (see $\S$~3.1.3) in a given band
  and test field. The residual fake columns are the percentage of all
  sources that are selected by these criteria that are still
  fake/spurious.}
\end{deluxetable}

\subsubsection{$cc\_flags$}
The confusion and contamination flag ($cc\_flags$) parameter indicates
if a source is a likely diffraction spike (D), scattered light halo
(H), image latent (P) or optical ghost image artifact
(O). Table~\ref{tab:ccflag} lists the fraction of sources by band and
field that have any of the flags `D', `H', `O' or `P' in that
band. These flags show that only a small and comparable percentage of
real and fake sources are flagged as any one of these image
contaminants in these fields.

\begin{deluxetable}{ccccc}
\tablewidth{0pt} 
\tabletypesize{\scriptsize}
\tablecaption{$cc\_flags$}
\tablehead{\colhead{} & \multicolumn{2}{c}{Real} & \multicolumn{2}{c}{Fake} \\
\colhead{Band} & \colhead{W3 \% DHO or P} & \colhead{W5 \% DHO or P} & \colhead{W3 \% DHO or P} & \colhead{W5 \% DHO or P} }
\startdata
1 & 2 & 1 & 1 & 1 \\
2 & 6 & 1 & 9 & 5\\
3 & 3 & 1 & 4 & 1 \\
4 & 8 & 2 & 13 & 5
\label{tab:ccflag}
\enddata 
\tablecomments{Data show the percentage of either `truth'
  catalog sources or fake sources in the {\it AllWISE} catalog that
  have any of the confusion and contamination flags D, H, O or P in a
  given field and band.}
\end{deluxetable}

\subsubsection{Profile Fit versus Aperture Magnitudes}
The {\it AllWISE} catalog provides magnitudes calculated from the
profile fit process ($w?mpro$) and also from curve-of-growth corrected
aperture photometry ($w?mag$). Real and fake sources show different
distributions in the difference between the profile fit and aperture
magnitude values as shown in the Appendix in
Figure~\ref{fig:psf-ap}. In Table~\ref{tab:psf-ap} we present the
retrieval rates and residual contamination using a simple trial cut,
requiring: $|w?mag - w?mpro| < 0.5$. Figure~\ref{fig:psf-ap} and
Table~\ref{tab:psf-ap} show that in bands 1 and 2 real and fake
sources are more easily separated by a cut based on the difference
between aperture and profile fit magnitudes. However in bands 3 and 4
the large spread in $|w?mag - w?mpro|$ of real sources and the large
quantity of fake sources with $|w?mag - w?mpro| \sim 0.5$ makes such
an approach uneffective.

\begin{deluxetable}{ccccc}
\tablewidth{0pt} 
\tabletypesize{\scriptsize}
\tablecaption{Profile Fit and Aperture Magnitudes}
\tablehead{\colhead{} & \multicolumn{2}{c}{Real Retrieval Rate} & \multicolumn{2}{c}{Residual Fake (\% of total)} \\
\colhead{Band} & \colhead{W3} & \colhead{W5} & \colhead{W3} & \colhead{W5} }
\startdata
1 & 61.5 & 74.3 & 3.8 & 1.5 \\
2 & 62.5 & 73.2 & 9.5 & 5.1 \\
3 & 45.7 & 51.2 & 65.9 & 63.6 \\
4 & 63.7 & 55.5 & 76.3 & 77.9 
\label{tab:psf-ap}
\enddata
\tablecomments{Retrieval rates are given as the percentage of
  real `truth' catalog sources that are retrieved by the selection
  criteria in $\S$~3.1.5 in a given band and test field. The residual
  fake columns are the percentage of all the sources that are selected
  by these criteria that are still fake/spurious.}
\end{deluxetable}

\subsection{An Approach to Mitigating Contamination} \label{sec:mitig}
\begin{figure}
\centering
\includegraphics[width=8cm]{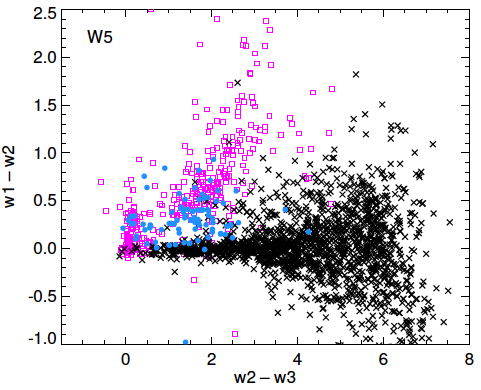}
\caption{{\it WISE} color-color diagram. Black $\times$-points are
  catalog entries with fake $w3$
  photometry.\label{fig:color-real-fake} Magenta points are Taurus
  YSOs from Rebull et~al.\ (2010) and blue points are transition disk
  sources from Andrews et~al.\ (2011) and Cieza et~al.\ (2012) found
  in the {\it AllWISE} catalog.}
\end{figure}

In this paper, our goal is to search for young stars with infrared
excess emission which, given the techniques laid out in K12, requires
that we obtain reliable photometry in at least 3 bands. The key
parameters that appear to have the strongest discriminatory effect in
suppressing fake source contamination in {\it WISE} bands $w3$ and
$w4$ are SNR and $\chi_\nu^2$. While combining many parameters to
reduce fake sources as much as possible is appealing, the cost would
be to sacrifice real sources we are interested in. Our approach is
thus to mitigate contamination using only the criteria based on signal
to noise and profile-fit reduced chi-squared. As noted in
$\S$~\ref{sec:snrchi}, the retrieval rate of real sources in bands 3
and 4 is poor using these criteria, so we recover low signal to noise
sources in band 3 by allowing sources with $0.45 <$ $w3rchi2 <
1.15$ and $w3snr > 5$, which means our potential retrieval rate of
real sources rises to about 60\% in $w3$. Although such a decision
also brings in a large quantity of fake detections, in our particular
application, the observed {\it WISE} colors of young stars provide an
additional discriminant for these objects. We use a 2$\arcsec$ search
radius to find YSOs in the {\it AllWISE} catalog from the Taurus
compilation of \citet{rebull10} and the transition disk objects (TDs)
in \citet{andrews11} and \citet{cieza12}. Transition disks are a
subclass of young star with inner opacity holes in their disks and
reduced levels of near infrared and/or mid-infrared excess
emission. The color-color diagram shown in
Figure~\ref{fig:color-real-fake} shows the distinct locations occupied
by these known YSOs and catalog entries with fake $w3$ photometry in
our test fields. Thus in searching for YSOs in {\it WISE} color-space
we can avoid regions with a high probability of reliance on a spurious
catalog entry.

The uncertainty/signal-to-noise/chi-squared criteria we use in the
following work are as follows. The color-magnitude criteria for YSOs
follow in $\S$~\ref{sec:scheme}.

{\it WISE} band 1: non-null $w1sigmpro$ and $w1rchi2 < (w1snr - 3)/7$.

{\it WISE} band 2: non-null $w2sigmpro$.

{\it WISE} band 3: $w3snr \geq 5$ and either: $w3rchi2 < (w3snr - 8)/8$ OR
$0.45 < w3rchi2 < 1.15$ (the approximate 1$\sigma$ range of
real-source $w3rchi2$ below the initial chi-squared-signal to noise
cut).

{\it WISE} band 4: non-null $w4sigmpro$ and $w4rchi2 < (2 \times w4snr - 20)/10$.

\section{The YSO Finding Scheme} \label{sec:scheme}
\begin{figure}
\centering
\includegraphics[width=12cm]{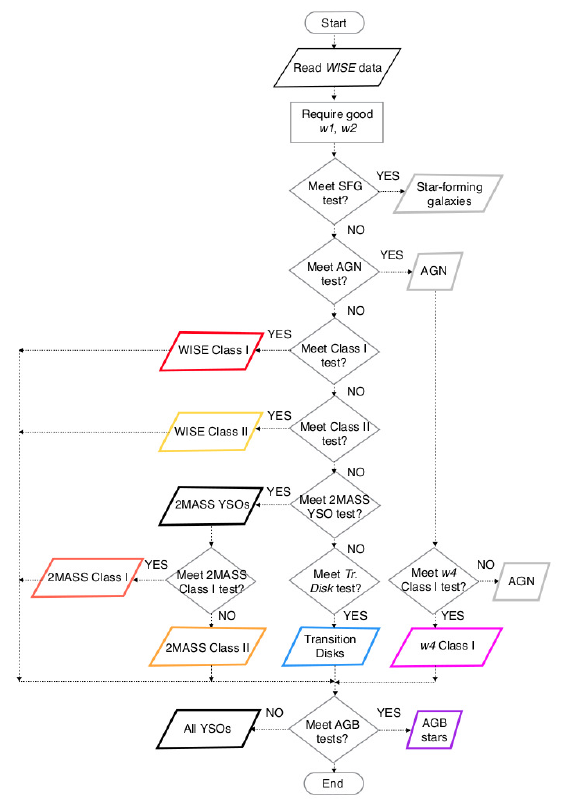}
\caption{Summary flowchart of the YSO identification and
  classification process in this paper.}
\label{fig:flowchart}
\end{figure}

With the previously described requirements on catalog parameters in
hand, we lay out the YSO identification and classification scheme. The
scheme resembles the one described in K12 but contains updates
designed to better account for and eliminate potential photometric and
astrophysical contaminants. Figure~\ref{fig:flowchart} summarizes the
entire scheme in a flow chart for reference. The photometric quality
criteria for {\it WISE} bands 1 and 2 are applied at all steps of the
scheme and are shown in a separate, preliminary box before any of the
classification steps are taken. The other band requirements are used
as needed in the scheme as described in the following text.

\subsection{Astrophysical Extragalactic Contaminants}
\begin{figure}
\centering
\includegraphics[width=15cm]{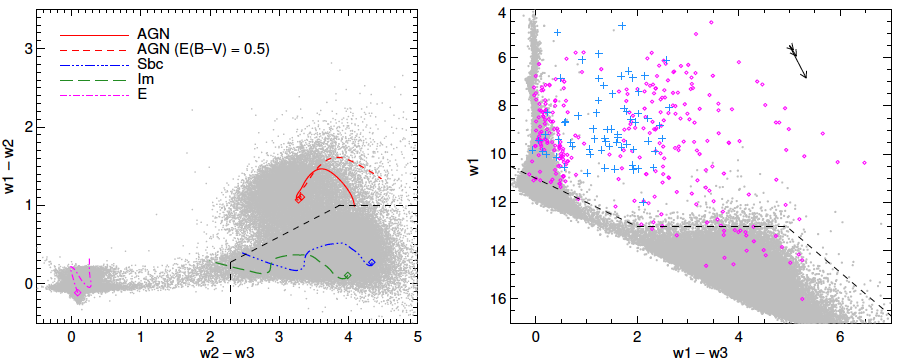}
\caption{Color-color and color-magnitude diagrams of {\it AllWISE}
  sources found the North Galactic Pole (gray
  points).\label{fig:galaxies} Left panel: we overlay theoretical SED
  template tracks of galaxy components from Assef
  et~al.\ (2010). Right Panel: Magenta points are Taurus YSOs from
  Rebull et~al.\ (2010) and blue points are transition disk sources
  from Andrews et~al.\ (2011) and Cieza et~al.\ (2012). The dashed
  black lines mark the regions excluded from our YSO selection. Arrows
  show extinction vectors of $A_{KS}$ = 0.4, 0.8 and 2.}
\end{figure}

We first remove objects from the catalog that are likely to be
unresolved external galaxies. The requirements for their removal are
similar to those of K12, but in this case based on sources found in
the {\it AllWISE} catalog around the North and South Galactic Poles,
which will likely consist mostly of galaxies and Galactic halo
stars. We extract sources from circles of area 100 square degrees
centered on the Poles by querying {\it AllWISE} for sources with $|b|
> 84.3558\degr$. Figure~\ref{fig:galaxies} shows a color-color diagram
of objects detected in the North polar field (gray points). We
overplot tracks showing the location of the galaxy SED templates of
\citet{assef10}. These templates are intended to reproduce the typical
components of galaxies: old (E/`elliptical'), intermediate
(Sbc/`spiral'), and young stellar populations (Im/`irregular'), as
well as nuclear activity (AGN, i.e. active galactic nuclei). Since
these selection cuts rely on {\it WISE} bands 1, 2 and 3, we
additionally apply the $w3$ photometry requirement of
$\S$~\ref{sec:mitig} and remove objects meeting all the following
criteria as likely star-forming galaxies (`SFG' in
Fig.~\ref{fig:flowchart} and the region marked in the left panel of
Fig.~\ref{fig:galaxies}):

\begin{eqnarray}
w2-w3 & > & 2.3 \\ 
w1-w2 & < & 1.0 \\
w1-w2 & < & 0.46 \times (w2-w3) - 0.78 \\
w1 & > & 13.0
\end{eqnarray}

Likely broad-line AGN are removed with a color-magnitude cut, see
Figure~\ref{fig:galaxies} (right panel). We label as candidate AGN
those sources that meet the band 1, 2 and 3 requirements and either:

\begin{eqnarray}
w1 & > & 1.8 \times (w1-w3) + 4.1 \\
\textrm{and} \nonumber\\
w1 & > & 13.0  \\ 
\textrm{or } \nonumber\\
w1 & > & w1-w3 + 11.0
\end{eqnarray}

In these and the following figures, we include arrows representing the
extinction vector for $A_{KS}$ = 0.4, 0.8 and 2, computed using an
interpolation of Figure 1 of \citet{mcclure09}. We present our
estimated extinction law in the {\it WISE} bands in
Table~\ref{tab:ext-law}. The vectors shown in the various figures are
not parallel to each other due to the variations in the extinction law
as a function of the total extinction as derived by \citet{mcclure09}.

\begin{deluxetable}{cccc}
\tablewidth{0pt} 
\tabletypesize{\scriptsize}
\tablecaption{$WISE$ Extinction Law}
\tablehead{\colhead{Band} & \colhead{$A_{\lambda}/A_{KS} (A_{KS}\leq0.5)$} & \colhead{$A_{\lambda}/A_{KS} (0.5< A_{KS}\leq1)$} & \colhead{$A_{\lambda}/A_{KS} (A_{KS}>1)$} }
\startdata
1 & 0.44 & 0.54 & 0.67 \\
2 & 0.26 & 0.344 & 0.51 \\
3 & 0.248 & 0.37 & 0.506 \\
4 & 0.155 & 0.27 & 0.43 
\label{tab:ext-law}
\enddata
\end{deluxetable}

\subsection{Young Stellar Objects} \label{sec:yso}
\begin{figure}
\centering
\includegraphics[width=16cm]{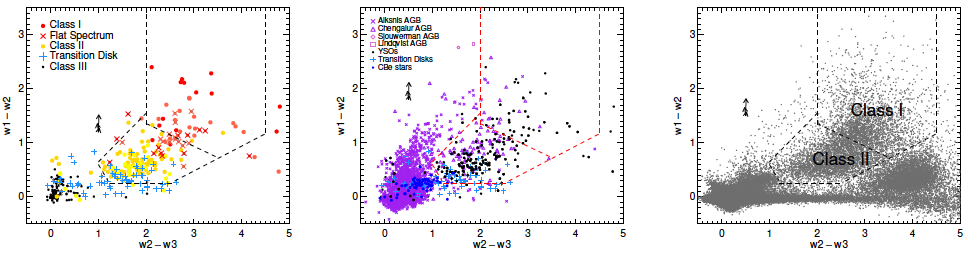}
\caption{{\it WISE} color-color diagrams. Left panel: Taurus YSOs from
  Rebull et~al.\ (2010) found in {\it AllWISE} and blue crosses:
  transition disks from Cieza et~al.\ (2012).\label{fig:ClI_II} Center
  panel: AGB stars, CBe stars, YSOs and transition disks, see
  text. Right panel: Outer Galaxy point sources from {\it AllWISE}
  found in region defined by $105 < l< 153, |b| < 5$. All objects
  shown meet the $w1$, 2 and 3 requirements of
  $\S$~\ref{sec:mitig}. Dashed lines show our YSO class
  divisions. Arrows show extinction vectors of $A_{KS}$ = 0.4, 0.8 and
  2.}
\end{figure}

Having removed the previously defined contaminants, we identify and
classify YSOs first using the list of objects that meet the band 1, 2
and 3 requirements. The scheme---as in K12---is developed to follow
the observed colors of Taurus region YSOs in \citet{rebull10} and
transition disk young stars, guided by the catalogs of
\citet{andrews11} and the compilation of \citet{cieza12}. To better
understand the distribution in {\it WISE} color-color and
color-magnitude spaces of potential astrophysical contaminants to our
young star candidates, we use the AGB carbon star catalog of
\citet{alks}, the OH/IR AGB star catalogs of \citet{chenga},
\citet{sjouw} and \citet{lindq} and the Classical Be stars from
\citet{mathew}. We search the {\it AllWISE} catalog within a
2$\arcsec$ search radius for all these object types, requiring their
catalog parameters pass our signal to noise and $\chi^2$ cuts and plot
them in Figure~\ref{fig:ClI_II} (young stars in the left panel and
other object types in the center panel). To guide the large scale
classification of AllWISE data, we also extract a large section of the
Outer Galaxy (105 $< l <$ 153, $|b| < 5$, hereafter: the Outer Galaxy
strip) to examine how these different object types manifest themselves
in the catalog (Fig.~\ref{fig:ClI_II} right panel). In contrast to
Fig.~\ref{fig:galaxies}, several additional clouds of sources can now
be seen in the right panel of Fig.~\ref{fig:ClI_II}. The extragalactic
objects are still obvious and present, but are now joined by Galactic
young stars and the other Galactic object types with infrared excess
seen in the left and center panels of the figure. The streak of
objects with roughly zero $w1-w2$ color but a wide range of $w2-w3$
color are a mix of source types. They may be objects with fake band 3
detections that slip through the photometric quality criteria of
$\S$~\ref{sec:mitig}. They may also be foreground photospheres/main
sequence stars with a coincidental background galaxy that gives them
an apparent red color. Finally, some fraction may be young transition
disk stars, or older, evolved `debris disk' stars similar to Vega
\citep[see for example,][]{dominik}.

Class I YSOs (candidate protostars) are the reddest objects, and are
classified as such if their colors match:

\begin{eqnarray}
w2-w3 & > & 2.0 \\ 
\textrm{and} \nonumber\\ 
w1-w2 & > & -0.42 \times (w2-w3) + 2.2 \\
\textrm{and} \nonumber\\ 
w1-w2 & > & 0.46 \times (w2-w3) - 0.9\\ 
\textrm{and}\nonumber\\
w2-w3 & < & 4.5
\end{eqnarray}

These requirements modify the scheme of K12 to better reflect the
divisions in SED slope ($\alpha$ = $d \log(\lambda F_\lambda)/d \log
\lambda$, Greene et~al.\ 1994) on which the Class system is based and
to eliminate contamination by spurious detections in {\it WISE} band
3.

Class II YSOs (candidate T Tauri stars and Herbig AeBe stars) are
classified from the remaining pool of objects if their colors match
the following criteria:

\begin{eqnarray}
w1-w2 & > & 0.25\\
\textrm{and} \nonumber\\
w1-w2 & < & 0.9 \times (w2-w3) - 0.25\\
\textrm{and} \nonumber\\
w1-w2 & > & -1.5 \times (w2-w3) + 2.1\\
\textrm{and} \nonumber\\
w1-w2 & > & 0.46 \times (w2-w3) - 0.9\\
\textrm{and}\nonumber\\
w2-w3 & < & 4.5
\end{eqnarray}

As can be seen in Figure~\ref{fig:ClI_II}, the revised Class II
criteria attempt to avoid the AGB stars seen in the catalog of Alksnis
et al. and the region occupied by Classical Be stars to the blue. This
latter region is also where some AGB stars reside, as well as Class
III/weak disk YSOs and blue transition disks \citep{cieza12}. The
criteria also attempt to eliminate a region likely dominated by the
reddest star forming galaxies to the lower right in this color-space.

\subsubsection{YSOs found with {\it 2MASS} and {\it WISE}}
\begin{figure}
\centering
\includegraphics[width=13cm]{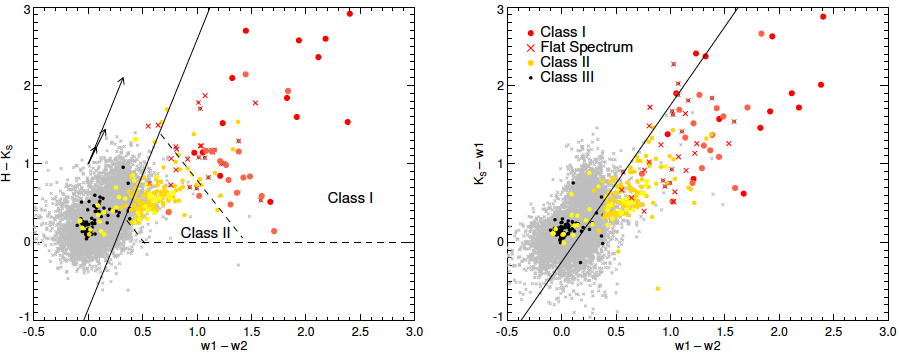}
\caption{{\it 2MASS-WISE} color-color diagrams. Gray points: point
  sources around the North Galactic Pole.\label{fig:2masswise} Colored
  points: YSOs from Taurus (Rebull et~al.\ 2010). Arrows show
  extinction vectors of $A_{KS}$ = 0.4, 0.8 and 2. Solid black line
  parallels the extinction vector for the highest range of extinctions
  tabulated in Table~\ref{tab:ext-law}. Dashed lines in left panel
  show the YSO Class divisions.}
\end{figure}

To compensate for the relative drop in sensitivity in {\it WISE} bands
3 and 4, both intrinsic and due to the bright background emission at
these wavelengths, we use the {\it 2MASS} $JHK_S$ point-source catalog
\citep{skrut06} that is automatically provided with the {\it AllWISE}
catalog. Here we alter the scheme from K12 and use the $H-K_S$
vs. $w1-w2$ color-color diagram to find and classify YSOs. In $H-K_S$
vs. $w1-w2$ color-color space there is less overlap between the
reddening vector due to extinction and that due to dust excess
emission (see Figure~\ref{fig:2masswise}). In other words, it is
easier to define a line that follows the standard extinction law that
excludes most extragalactic contamination for the $H-K_S$, $w1-w2$
plot than it is in the $K_S-w1$, $w1-w2$ plot, where reddened
background galaxies and main-sequence stars may easily have colors
resembling disk excess sources. 

One further alteration we make to the scheme of K12 and the original
{\it Spitzer-2MASS} concept of \citet{gutermu09} is to not apply any
dereddening to the colors of sources we classify. Lacking an
independent method to remove the foreground extinction to each source,
or an extinction map, we develop our selection and classification
based on the raw photometry. In general, in the Outer Galaxy, line of
sight extinction due to dust in between the Solar System and the main
star forming regions is low, for example \citet{hillwig} derive an
extinction $A_V\approx2$ to the O stars in the \htwo\ region cavity of
W5. Within star forming regions however, the most embedded young stars
can be hidden behind an $A_V>$15 \citep[for example,][who find peak
  $A_V=$19.8 in their extinction map of AFGL~490]{masiunas12}. Thus in
general, the extinction vector in $H-K_S$ versus $w1-w2$ will have a
magnitude $<0.3$, but could be as large or larger than 1 in the most
embedded cases, meaning some of the objects classified as Class I
protostars in this part of the scheme could be reddened Class II
stars. 

We search for YSO candidates using the following criteria amongst
previously unclassified objects with non-null photometric error in
{\it 2MASS} $H$ and $K_S$, and which pass the {\it WISE} band 1 and 2
requirements of $\S$~\ref{sec:mitig} and {\it fail} those for
{\it WISE} band 3:

\begin{eqnarray}
H-K_S & > & 0.0 \\
\textrm{and }\nonumber\\
H-K_S & > & -1.76 \times (w1-w2) + 0.9\\
\textrm{and }\nonumber\\
H-K_S & < & (0.55/0.16) \times (w1-w2) - 0.85 \\
\textrm{and }\nonumber\\
w1 & \leq & 13.0
\end{eqnarray}

These objects are initially classified with the Class II objects. The
subset of them that pass the following criteria are grouped with the
candidate Class I/protostar objects:

\begin{equation}
H-K_S > -1.76 \times (w1-w2) + 2.55\\
\end{equation}

The ratio (0.55/0.16) in the criteria above comes from the calculation
of the ratio of color-excesses: $(A_H - A_{KS})/(A_{w1} - A_{w2})$
computed for $A_{KS}>1$, since this gives the greatest slope in the
extinction law, and thus the most cautious cut of galaxies in $H-K_S$,
$w1-w2$ space.

\subsubsection{{\it WISE} 22~$\micron$ Photometry}
\begin{figure}
\centering
\includegraphics[width=16cm]{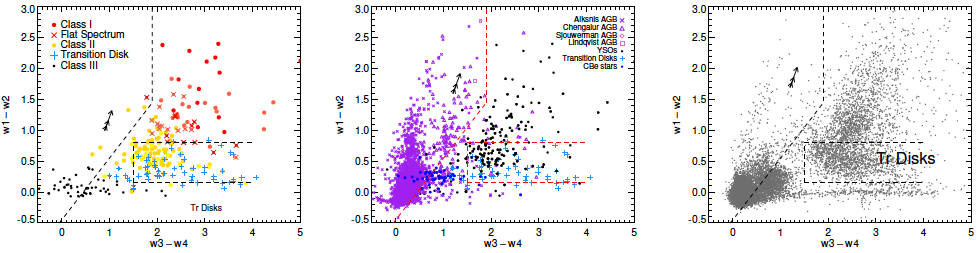}
\caption{{\it WISE} 4-band color-color diagrams. Left panel: Taurus
  YSOs from Rebull et~al.\ (2010) found in {\it AllWISE} and blue
  crosses: transition disks from Cieza
  et~al.\ (2012).\label{fig:tds-agb} Center panel: AGB stars, CBe
  stars, YSOs and transition disks colored as in
  Figure~\ref{fig:ClI_II}. Right panel: Outer Galaxy point sources
  from {\it AllWISE} found in region defined by $105 < l < 153, |b| <
  5.$ All objects shown meet the requirements on all four {\it WISE}
  bands in $\S$~\ref{sec:mitig}. Dashed lines show the upper left
  region excluded to remove OH/IR AGB stars and the color-criteria for
  finding transition disk candidates. Arrows show extinction vectors
  of $A_{KS}$ = 0.4, 0.8 and 2.}
\end{figure}

As in K12, we use $w4$ photometry to identify candidate transition
disk objects and to retrieve possible protostars from the AGN
candidates. Figure~\ref{fig:tds-agb} shows the location of the search
box for transition disk objects. As can be seen, there is considerable
overlap between literature transition disks and Class II sources,
likely owing to the sensitive dependence of mid-infrared emission on
small changes in the dust distribution and disk inclination. These
properties make a scheme like this one an imprecise tool for finding
these objects. The streak of objects with roughly zero $w1-w2$ color
and a range of $w3-w4$ colors is similar to the one seen in
Fig.~\ref{fig:ClI_II} and is likely made up of the same types of
sources. As can be seen, literature transition disks are typically not
as blue in $w1-w2$ as this streak, so we avoid this portion of
color-color space. We select candidate transition disks with the
following criteria, from the sources that meet the requirements on all
four bands given in $\S$~\ref{sec:mitig}.

\begin{eqnarray}
w3-w4 & > & 1.5\\
\textrm{and}\nonumber\\
0.15 & < & w1-w2 < 0.8\\
\textrm{and}\nonumber\\
w1-w2 & > & 0.46 \times (w2-w3) - 0.9\\
\textrm{and}\nonumber\\
w1 & \leq & 13.0
\end{eqnarray}

The last two criteria attempt to mitigate the contamination by
background galaxies in the same way as with Class II objects. The
bluest transition disk objects seen in the literature catalogs are too
hard to separate from AGB and CBe stars without follow-up spectroscopy
\citep[as in][]{cieza12} and we make no attempt to find them in this
work.

\begin{figure}
\centering
\includegraphics[width=8cm]{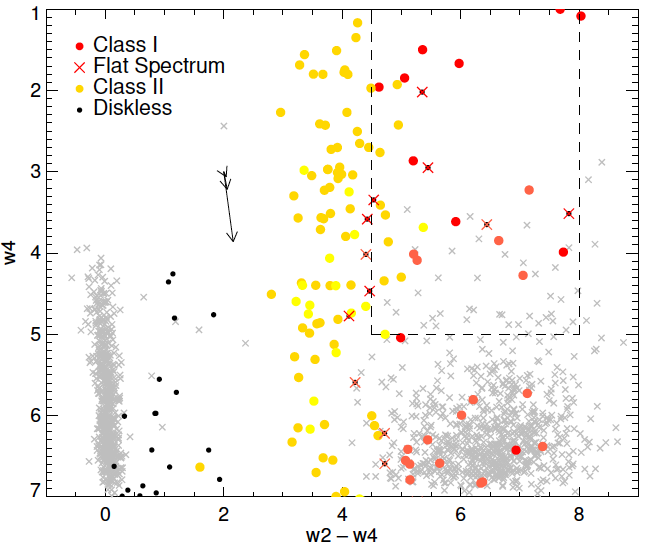}
\caption{{\it WISE} color-magnitude diagram. Gray points: point
  sources around the North Galactic Pole.\label{fig:AGN-retrieve}
  Colored points: YSOs from Taurus (Rebull et~al.\ 2010), colored as
  before. Arrows show extinction vectors of $A_{KS}$ = 0.4, 0.8 and
  2. Dashed box shows region searched for protostars retrieved from
  AGN sample.}
\end{figure}

AGN occupy a similar part of color space to protostars, but we can
retrieve an additional set of those objects as candidate protostars if
they have a $w4$ magnitude that makes them brighter than the majority
of external galaxies. We take the sample of objects previously
classified as AGN candidates and identify new protostars with the
following color-magnitude cuts, additionally requiring that they meet
the band 4 quality criteria given in $\S$~\ref{sec:mitig} (see
Figure~\ref{fig:AGN-retrieve}):

\begin{eqnarray}
w4 & < & 5.0 \\
\textrm{and}\nonumber\\
4.5 & < & w2-w4 < 8.0\\
\textrm{and}\nonumber\\
w1-w2 & > & 1.0\\
\textrm{and}\nonumber\\
w3-w4 & > & 2.0
\end{eqnarray}

\subsubsection{AGB stars and CBe stars}
\begin{figure}
\centering
\includegraphics[width=14cm]{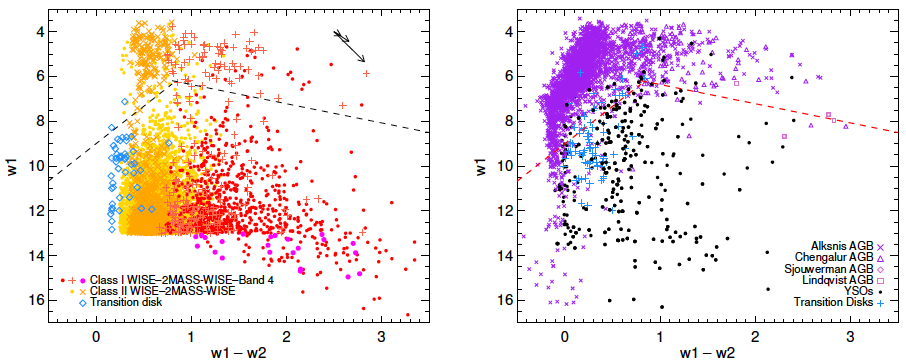}
\caption{{\it WISE} color-magnitude diagram.\label{fig:agb-cut} Left
  panel: {\it WISE} YSO candidates in the Outer Galaxy strip from
  $\S$~\ref{sec:yso}, classified using our scheme. Arrows show
  extinction vectors of $A_{KS}$ = 0.4, 0.8 and 2. Right panel:
  literature AGB stars, CBe stars, Taurus YSOs and transition
  disks. All objects are filtered using our $w1$ and 2
  criteria. Dashed lines shows boundary above which we eliminate
  sources as likely AGB stars.}
\end{figure}

Both AGB stars and CBe stars possess a degree of infrared excess,
originating in the circumstellar material around these objects. In the
case of AGB stars, these objects are in general much brighter than
YSOs, unless they are at great distance or strongly extincted, and
have a slightly different distribution in color-color space, with the
OH/IR category of AGB stars the most similar to YSOs \citep[a similar
  finding to][using {\it Spitzer} data]{robitaille08}. CBe stars in
general possess a weaker infrared excess than do YSOs, but can be
confused with blue transition disk objects. Our initial {\it WISE}
band 1, 2 and 3 classification of YSOs filters some fraction of these
object types. After passing through all the preceding steps, we make
one final pass to remove these objects. We use a combination of the
{\it WISE} $w1-w2$, $w3-w4$ color-color diagram (see
Figure~\ref{fig:tds-agb}) and the $w1$ vs. $w1-w2$ color-magnitude
diagram (Figure~\ref{fig:agb-cut}). An initial run of the
classification scheme up to this point on the Outer Galaxy strip
described in $\S$~\ref{sec:yso} is shown in the left panel of
Figure~\ref{fig:agb-cut}. The prominent clustering of objects with
bright $w1$ magnitudes is almost certainly dominated by AGB stars as
can be seen in the right panel of the figure. Thus, stars previously
classified as any type of YSO are rejected if they meet the following
conditions:

Firstly, any of the YSO candidates from the previous sections that are
bright in {\it WISE} band 1 with:

\begin{eqnarray}
w1 & < &  (-10/3) \times (w1-w2) + 9\\
\textrm{or}\nonumber\\
w1 & < &  (6/7) \times (w1-w2) + 5.5
\end{eqnarray}

or any of the YSO candidates from the previous sections that meet the
requirements on all four bands given in $\S$~\ref{sec:mitig} and have:

\begin{eqnarray}
w3-w4 & < & 1.9\\
\textrm{and }\nonumber\\
w1-w2 & < &  w3-w4 - 0.45
\end{eqnarray}

The right panel of Figure~\ref{fig:agb-cut} also shows that for nearby
regions like Taurus, some bright YSOs may be lost to a cut for AGB
stars like this one. In cases like this, analysis of the data for the
region considered would be necessary to establish a compromise cut
appropriate for the data considered.

\subsection{Planetary Nebulae}
\begin{figure}
\centering
\includegraphics[width=14cm]{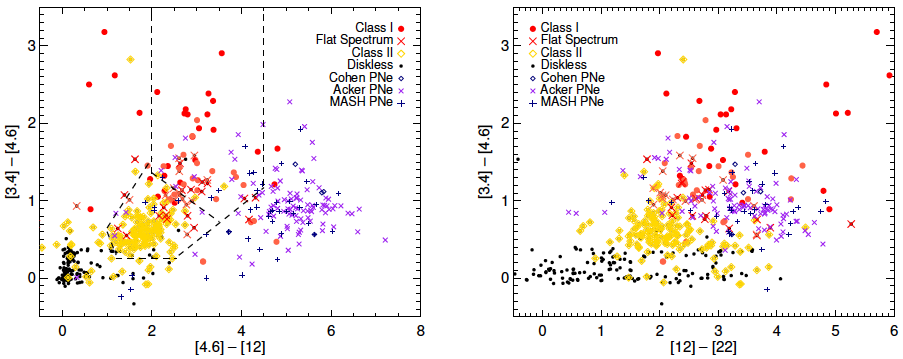}
\caption{{\it WISE} color-color diagrams showing YSOs from Taurus
  (Rebull et~ al.\ 2010) and PNe from Parker et~al.\ (2006), Cohen
  et~al.\ (2011) and Acker et~al.\ (1992).\label{fig:PNe}}
\end{figure}

Unresolved planetary nebulae (PNe) are potentially a source of
contamination to the YSO sample, and are considered in
$\S$~\ref{sec:resid} when we compute the residual contamination likely
from Galactic sources. However, we make no attempt to avoid their
detection in this scheme. As can be seen in Figure~\ref{fig:PNe},
their typical colors in the {\it WISE} bands when filtered by our
signal-to-noise and chi-squared cuts place them mainly outside the
regions occupied by young stars and our selection regions. We draw
samples of PNe from a 2$\arcsec$ search in {\it AllWISE} for objects
in the Macquarie/AAO/Strasbourg H$\alpha$ (MASH) catalog
\citep[][]{parker06,miszalski08}, \citet{cohen11} and \citet{acker92}.

A recent paper by \citet{nikutta} used a model of a spherical
circumstellar dust shell to predict the location of AGB stars and
young stars in {\it WISE} color-color diagrams. This model leads them
to conclude that young stars should be found in the streak of objects
that we note in sections 4.2 and 4.2.2, which we suspect are more
likely to be stars with debris disks \citep[see, for
  example][]{morales}. Their model also leads them to predict that
young stars could be found by searching for objects with 10$<w1<$11,
0.4$<w1-w2 <$1.4, and 4$< w2-w3 <$5.7. As can be seen in
Figure~\ref{fig:PNe}, while some Class I young stars will be found by
such a search, the majority of this search space is occupied by
planetary nebulae.

\subsection{Residual Contamination by Fake Photometry}
Although the photometric quality requirements in $\S$~\ref{sec:mitig}
attempt to ensure a reliable source catalog is used to search for
young stars, it is possible that at the end of the process of cleaning
the {\it AllWISE} catalog and classifying the sources that some of the
YSO candidates are reliant on a spurious detection for their
classification.

We test the combined effect of the photometric quality criteria and
the YSO identification scheme on our W3 and W5 test fields by
classifying the {\it AllWISE} catalog sources and flagging those where
the classification was reliant on a fake detection. We find 1/11 {\it
  WISE} Class I's (9\%), and 1/76 {\it WISE} Class II's (1.3\%) had a
spurious $w3$ detection. None of the YSO candidates relied on a
spurious band 1 or 2 detection.

\subsection{Application of the Scheme to the Outer Galaxy Strip}
\begin{figure}
\centering
\includegraphics[width=15cm]{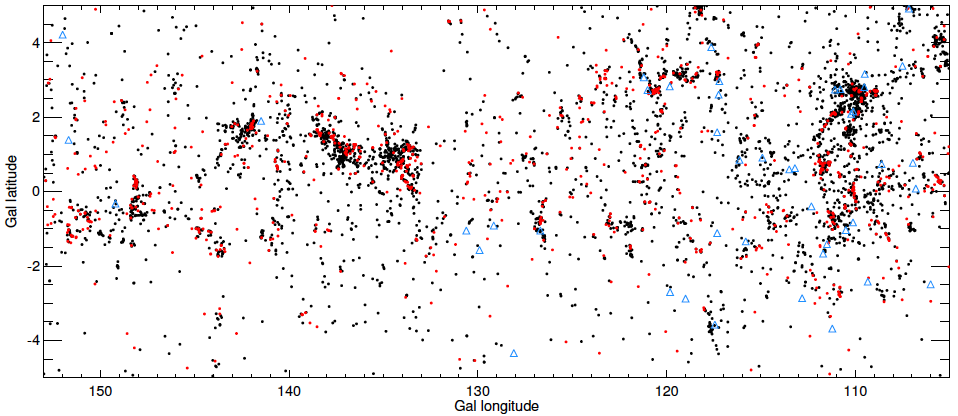}
\caption{Distribution of {\it AllWISE} detected and classified YSO
  candidates in the Outer Galaxy.\label{fig:galstrip} Red points:
  Class I sources, black points: Class II sources, blue triangles:
  transition disk candidates.}
\end{figure}

Having developed the classification scheme, we test it on the Outer
Galaxy strip described in $\S$~\ref{sec:yso}. We query the {\it
  AllWISE} online database, applying the criteria for $w1$ and $w2$ in
the built-in SQL tool, since these apply to all parts of the YSO
classification scheme. From an initial 9.1 million catalog sources, we
retrieve the following distribution of YSO types shown in
Figure~\ref{fig:galstrip} and listed in Table~\ref{tab:galstrip_ysos}.

\begin{deluxetable}{cccccc}
\tablewidth{0pt} 
\tabletypesize{\scriptsize}
\tablecaption{Galactic Strip YSOs}
\tablehead{\multicolumn{2}{c}{Class I} & \multicolumn{2}{c}{Class II} & \colhead{Transition Disk} & \colhead{Embedded Protostars} \\
\colhead{$WISE$} & \colhead{$2MASS WISE$} & \colhead{$WISE$} & \colhead{$2MASS WISE$} & \colhead{ } & \colhead{ }}
\startdata
750 & 148 & 3410 & 773 & 45 & 23 
\label{tab:galstrip_ysos}
\enddata
\end{deluxetable}

Several notable star forming regions are picked out by the
distributions of young stars, in particular (from left to right):
AFGL~490, the W5-W4-W3 complex, Cep~OB3 (Allen et al. 2012) and the
NGC~7538 complex (Ungerechts et al. 2000). The less evolved Class I
sources are fewer in number and appear to be spatially concentrated in
smaller structures than their neighboring Class II counterparts. We
return to this point in a later section.

\subsection{Residual Contamination by Galaxies and Galactic Objects} \label{sec:resid}
We note that despite the scheme, contaminant astrophysical objects may
contribute to the observed distribution of young stars in the Outer
Galaxy Strip seen in Fig.~\ref{fig:galstrip}. To estimate the number
of these contaminants we run the following tests.

For external galaxies we download an equivalent area of sky around the
North and South Galactic poles (simply searching the {\it AllWISE}
catalog with $|b| > 81.25668$). We randomly place these objects in the
Outer Galaxy strip and retrieve the extinction value of the nearest
pixel of the \citet{schlegel98} extinction map for each object. We
redden their {\it 2MASS} and {\it WISE} photometry according to the
extinction law in Table~\ref{tab:ext-law} and run the full catalog
cleaning and classification scheme. We present the average number of
objects that fall into our various YSO class bins in this particular
field after 20000 repeats of this process in
Table~\ref{tab:resid_gals}. As expected, {\it WISE}-selected Class I
YSOs are the most highly contaminated by galaxies at a rate of about
13\%. All other classes are contaminated by external galaxies at a
rate of 2\% or less.

\begin{deluxetable}{cccccc}
\tablewidth{0pt} 
\tabletypesize{\scriptsize}
\tablecaption{Probable Extragalactic Contamination in the Outer Galaxy Strip}
\tablehead{\multicolumn{2}{c}{Class I} & \multicolumn{2}{c}{Class II} & \colhead{Transition Disk} & \colhead{Embedded Protostars} \\
\colhead{$WISE$} & \colhead{$2MASS WISE$} & \colhead{$WISE$} & \colhead{$2MASS WISE$} & \colhead{ } & \colhead{ }}
\startdata
96.07 & 1.81 & 66.78 & 8.67 & 0.002 & $<5\times10^{-5}$
\label{tab:resid_gals}
\enddata 
\tablecomments{Figures in this table are averages of 20,000 simulation
  runs computing the number of contaminant extragalactic objects
  misclassified into the YSO classes presented in this paper. See
  $\S$~\ref{sec:resid} for details.}
\end{deluxetable}

For Galactic contaminants we use the Galactic populations model of
\citet{wainscoat92} and the extension to all infrared wavelengths of
\citet{cohen93} over this field of view, but with no extinction, so
the results are an upper limit (a code implementation of this model
was kindly supplied by Miranda Dunham and Paul Harvey). We select the
following categories of objects from Table 2 of Wainscoat et~al.\ as
likely contaminants for confusion with our YSOs in Class I or II by
their location in {\it WISE} or {\it 2MASS-WISE} color-color spaces:

Our Class I objects may be confused with Wainscoat types: X 1E, 2, 3,
4, (the ultraluminous 12~$\micron$ sources) AGB\_CI\_01--05, (carbon
rich, `optically invisible' AGB stars with IRAS $[12]-[25]$ color in
the range 0.1--0.5) AGB\_M\_17--25, (oxygen rich AGB stars with IRAS
$[12]-[25]$ color in the range 1.7--2.5) PNE BLUE, RED (`blue' and
`red' planetary nebulae).

Our Class II objects may be confused with Wainscoat types: X 1E, (the
ultraluminous 12~$\micron$ sources with silicate emission)
AGB\_CI\_01--31, (C-rich, `optically invisible' AGB stars,
0.1$<[12]-[25]<$3.1) AGB\_M\_07--15, (O-rich AGB stars,
0.7$<[12]-[25]<$1.5) RNE BLUE, RED (`blue' and `red' reflection
nebulae).

Since the Wainscoat-Cohen model code allows the user to select which
of the different object categories are present in a particular portion
of the Galaxy, we use it to generate $L$-band magnitude histograms for
these specific Class I and II contaminants separately, applying our
average {\it WISE} band 1 completeness curve for the W5 test field at
the faint end and cutting everything brighter than $L$=7 (to replicate
our AGB exclusion method). This method assumes that the $L$-band is a
good match for $w1$.

Objects in the X1E, AGB\_M\_07 and 09, AGB\_CI\_01--11 categories all
have colors which mean they would be rejected by our {\it WISE}
$w1-w2$ vs. $w3-w4$ category if detected in {\it WISE} band 4. We
remove these from our Galactic contaminant estimation if their
apparent IRAS 25~$\micron$ magnitude is less than 6 in the final
computation (an approximate band 4 completeness cut).

We run the Wainscoat-Cohen model code once for the Class I-like
sources and once for the Class II-like sources. We apply an additional
faint cut of 13th magnitude for the Class II sources, since these are
rejected by the YSO scheme, and derive a final numerical estimate of
the total number of non-YSOs that may be in the field of view. We
predict 68 Class I and 222 Class II sources at most may be contaminant
Galactic objects. Since we do not separate this simulated population
of Galactic contaminant objects into the different bins of our
classification scheme, these estimates represent $\sim$7.5\% and
$\sim$5\% of the total Class I and Class II YSO candidates found in
the Outer Galaxy strip. The robust population of young stars detected
by the {\it AllWISE} catalog and our classification scheme is thus
about 730 Class I and 3900 Class II objects.

\section{Analysis}
\subsection{Stellar Clustering}
The question of the existence of two distinct modes of star formation,
one `clustered' where stars are formed in groups, associations or true
clusters as observed in general in star forming regions, and one
`isolated' has been raised in the literature for decades. For example,
\citet{roberts57} attempted to calculate whether all O and B stars
form in clusters and associations, or whether ``individual stars form
in the general field.'' A bimodality in star formation clustering has
been suggested to explain, for example, the widespread detection of
X-ray bright and presumably, young stars in and around molecular
clouds \citep{carpenter00}. If an isolated population does exist, a
natural question would be to ask if it has different properties to the
clustered mode. In this paper, we use the term `isolated' or
`non-clustered' to specifically refer to a distinct population that is
separate from the continuum of cluster sizes that are observed in star
forming regions and molecular clouds. If a clustered environment is
necessary for the formation of massive stars (see Zinnecker \& Yorke
2007 for a discussion of this hypothesis), then the mass function of
the isolated population may differ from that of the clustered
population, although the calculations of \citet{dewit05} show that
isolated O stars can be explained by the combined effects of a
universal cluster size/density distribution and stellar mass function.

One of the simplest ways to analyze the clustering properties of stars
projected onto the sky is to compute the distribution of local surface
densities. We compute a proxy for the surface density, $\mu$ at the
position of each star using the angular distance to the 6th nearest
neighbor of each object, $s_6$. Following \citet{casertano85}, the
$j$th surface density estimator $\mu_j$ is:

\begin{equation}
\mu_j  =  \frac{j-1}{\pi s_j^2}
\end{equation}

where $s_j$ is the distance to the $j$th nearest neighbor of an
object. \citet{casertano85} conclude that choosing $j=6$ provides the
best balance between an accurate computation of density (which would
require large $j$) and a `local' density estimate, which would call
for small $j$. Thus:

\begin{equation}
s_6 \propto \mu_6^{-2}\\
\end{equation}

\begin{figure}
\centering
\includegraphics[width=13cm]{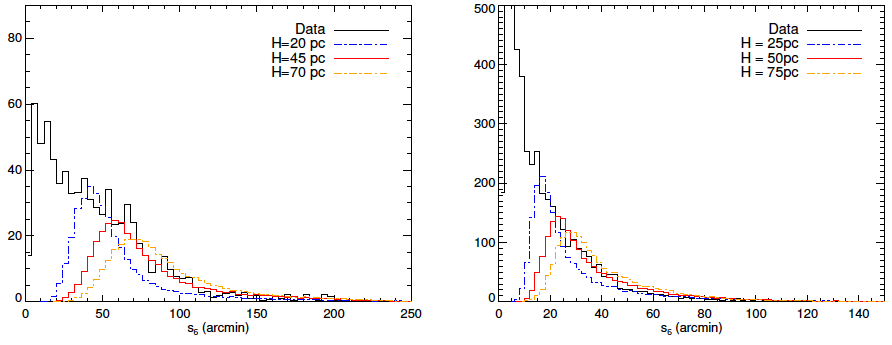}
\caption{Black histogram: 6th nearest neighbor distance $s_6$
  distribution for Class I candidates and combined Class I+II sample
  after subtracting off computed average distribution of galaxies and
  Galactic contaminants from Wainscoat et~al.\ (1992).\label{fig:AD}
  Colored histograms show the scale height distributed samples' $s_6$
  distributions for a range of trial scale heights.}
\end{figure}

Since we do not know exactly which of our YSO candidates are
background galaxies, or Galactic contaminant objects, we compute an
average angular distance distribution, applying a Monte Carlo style
method to subtract off these sources of contamination by repeatedly,
randomly removing samples of objects with the same $w1$ magnitude
distributions and total number as that predicted previously and
re-computing the angular distance distribution. In each iteration we
derive a new calculation of the expected contamination due to
galaxies. The final averaged angular distance distributions for the
Class I sources and for a combined sample of the Class I and II
sources are shown in Figure~\ref{fig:AD} in the left and right panels
respectively. Both distributions in Figure~\ref{fig:AD} peak at small
angular distances because of the large fractions of these objects in
clusters and aggregates of various sizes. Since our YSO survey is
incomplete, the true peaks of the distributions are likely at smaller
angular distances than those shown in the figure. The distributions
both have tails that extend to large angular distances $>$2$\degr$,
some of which could be explained by a population of very
isolated/non-clustered young stars.

We investigate the possibility of a non-clustered population by
assuming that such sources would exist in a distribution with some
scale height in the Galactic disk. For a simple test, we construct a
comparison sample with an exponential distribution in the vertical
(latitude) direction and uniform in the plane of the Galaxy. We assign
each object an absolute $w1$ magnitude by drawing from the
distribution of the c2d catalog sources of \citet{evans09}, where we
match their list to the {\it AllWISE} catalog and separate their
sources into the YSO Class divisions set in $\S$~\ref{sec:scheme}. We
apply the average W5 $w1$ completeness curve to the resultant apparent
magnitude distribution of sources and additionally require objects
have $w1 \leq 13$, since our extragalactic filtration technique
essentially imposes this cut on our data. In Figure~\ref{fig:AD}, we
show the angular distance distributions of these samples where we find
an approximate match to the shape of the low density end of the true
YSO distribution, separately matching Class I objects (left panel) and
a combined Class I and II sample (right panel). Increasing the total
number thrown into the sample (with constant scale height) pushes the
number of sources at the peak of the curve above the data. Decreasing
the total input number produces an excess of sources at very low
density/large angular distance (more than a factor of 2). The {\it
  WISE} YSO histograms contain on average 760 Class I and 3910 Class
II stars after contaminant subtraction. The best matching histograms
contain on average 320 sources (left panel) and 1800 sources (right
panel). If a uniform, scale height component makes up part of this
observed distribution of YSOs in this way, this analysis shows that it
is not more than $\approx$1/3 of the total number of sources.

We note that {\it WISE}'s incompleteness means the observed stellar
densities are underestimates (or in apparent nearest neighbor angular
distances, overestimates) of the true values. In the prominent star
forming regions in Figure~\ref{fig:galstrip} are the well-studied
regions: AFGL~490 (($l, b$) = (142.0, +1.82), distance = 900~pc,
Masiunas et~al. 2012), Cep OB3 (($l, b$) = (111.26, +2.98), $d$ =
700~pc, Allen et~al. 2012), W5 (($l, b$) = 138, +1.5, $d$ = 2~kpc,
Koenig et~al. 2008). A 3$\arcsec$ match of our YSO catalog to these
published lists of Class I and II sources (excluding transition disks
and Class III/weak-line T Tauri stars) retrieves 26\% (AFGL~490), 25\%
(Cep~OB3) and 16\% (W5) of the {\it Spitzer}-classified stars. Thus
the true stellar densities in these clusters are at least a factor
$\approx$5 greater.

\begin{figure}
\centering
\includegraphics[width=13cm]{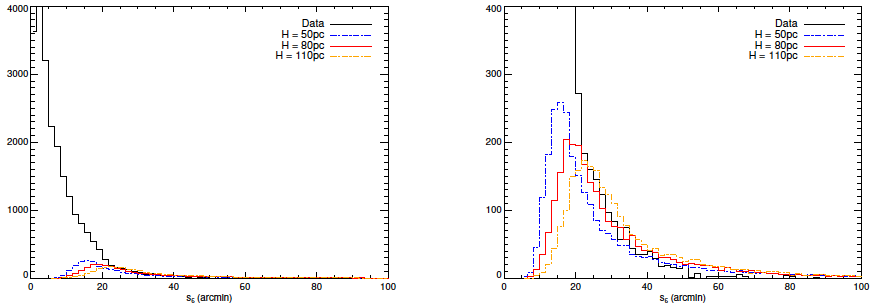}
\caption{Black histograms: 6th nearest neighbor distance $s_6$
  distribution for combined Class I+II samples scaling all distances
  by 1/$\sqrt{5}$ and increasing the total number by
  5.\label{fig:scaled-AD} Colored histograms show the scale height
  distribution sample $s_6$ distributions for a range of trial scale
  heights. Right panel is a zoom in on the left panel to highlight the
  differences in the model histograms.}
\end{figure}

To test if a scale height distributed component could still explain
the low density end of the young star density distribution, we crudely
imitate the effect of {\it WISE}'s incompleteness by scaling all the
nearest neighbor distances by 1/$\sqrt{5}$ and increasing the total
number of objects in the histogram by 5. This experiment assumes that
we have missed a fixed fraction of the total number of young stars in
the Outer Galaxy, whether in or out of massive regions. We repeat the
experiment of trying to match scale height distributions of stars to
the new, scaled histogram. In this test we apply the {\it Spitzer}
completeness calculation from \citet{koenig08} to the sample of
sources and apply a faint cut off of IRAC band 1 $< 14$ to represent
their more lenient galaxy removal
technique. Figure~\ref{fig:scaled-AD} shows the results of this trial
applied to the combined Class I and II sample. In this case, a smaller
fraction---roughly 2400 objects out of the increased YSO count of
23350 stars---can be fit with the uniform component, which also
suffers from the problem of over-predicting the number of low density
sources (nearest neighbor distance $> 50\arcmin$) by a factor of
5. Since {\it Spitzer} itself is also incomplete to the low mass end
of the IMF in this field, and we have only considered those young
stars which still possess a visible infrared excess, the fraction of
stars in a random, non-clustered population may be lower still.

\subsection{Two-point Correlation Function}
We compute the two point correlation function (TPCF) of a single
output of the contaminant-subtracted sample of YSOs, separated into
Class I and II samples. We use the following estimator of the TPCF
following \citet{peebles74}:

\begin{equation}
\xi(\theta) = \frac{N_R}{N}\frac{DD(\theta)}{RR(\theta)} - 1
\end{equation}

\begin{figure}
\centering
\includegraphics[width=15cm]{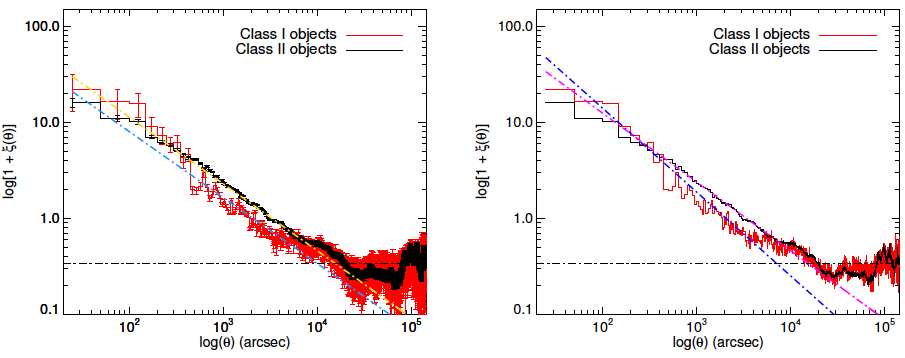}
\caption{Two-point correlation function for the Class I sources (red
  histogram) and Class II sources (black histogram). Left panel: dot
  dashed lines show weighted least-squares fits to the straight-line
  part of the data for Class I sources (blue) and Class II sources
  (yellow). In the right panel we remove the error bars and overplot
  unweighted least squares fits to the data between 200 and
  4000$\arcsec$ (Class I sources: blue and Class II sources:
  magenta).\label{fig:tpcf} Black dot-dashed line shows level of a
  random distribution in this formalism.}
\end{figure}

Where $DD$ and $RR$ are the numbers of pairs as a function of angular
separation $\theta$ in the YSO catalog and in a randomly distributed
catalog with 3$\times$ the number of objects respectively. We generate
$RR$($\theta$) by randomly populating the latitude-longitude range of
our test region. In this formulation, a random distribution of sources
would have $\xi$($\theta$) = $-2/3$, or $\log$(1 + $\xi$($\theta$)) =
$-0.477$. Figure~\ref{fig:tpcf} shows a plot of (1 + $\xi$($\theta$))
vs. $\theta$ for the Class I and II samples (with logarithmically
scaled axes), overlaid with weighted least-squares fits to the roughly
linear part of the data (left hand panel). The right hand panel shows
the same data without error bars and unweighted fits to the data
between 200 and 4000$\arcsec$.

Both the Class I and II data fall below the random level at
separations $\theta \gtrsim 18000\arcsec$, or about 5$\degr$, likely
due to edge effects becoming significant in the calculation of
$\xi$($\theta$). The features evident in the data at the longest
separations (roughly 100,000$\arcsec$ or 28$\degr$) are the result of
the real clusters visible in Figure~\ref{fig:galstrip}. The Class I
distribution appears to follow a power law up to separations of about
4000$\arcsec$ or just about one degree. The Class II distribution is
shallower, and appears to have a power law shape up to separations
$\theta \approx 7000\arcsec$ with a turnover at about 200$\arcsec$. As
discussed by \citet{kraus}, work by \citet{larson}, \citet{simon97}
and others has shown evidence that young stars are distributed in a
fractal manner with self-similar structure over a range of angular
scales in star forming regions like Taurus. Following the methodology
of Kraus \& Hillenbrand, in a fractal distribution the number of
neighbors within angular distance $\theta$ goes as N($\theta$)
$\propto$ $\theta ^{D}$, where $D$ is the fractal dimension, or the
degree to which the distribution fills space. In such a distribution
the mean surface density of companions $\Sigma$($\theta$) $\propto$ 1
+ $\xi$($\theta$) $\propto$ $\theta^{D-2}$. For the Class I sources
the weighted least squares fit technique finds a slope of
$-0.69\pm0.06$ and thus $D = 1.31$ and for the Class II's a slope of
$-0.70\pm0.02$ and $D = 1.30$. These slopes are statistically equal,
but the fits are strongly skewed to match the data around
2000$\arcsec$ where the error bars are smallest, due to the random
distribution peaking around this value of object separations. The
unweighted fits (slopes of $-0.87\pm0.02$ for the Class I's and
$-0.72\pm0.01$ for the Class II's) capture the fact the Class I source
correlation function falls below the Class II function at scales
$>300\arcsec$, consistent with the more evolved Class II sources being
more spread out in space and the younger Class I sources more
filamentary and concentrated. This interpretation is consistent with
the findings of \citet{andre10} and others who show that star-forming
gas as seen in the far-infrared with the {\it Herschel} space
telescope appears to be dominated by filamentary structure. The
difference in the intercepts of the straight-line fits in the
left-hand panel of Figure~\ref{fig:tpcf} is $\approx$0.17, or a size
ratio of about 1.5.

\subsection{Cluster Dispersal} \label{sec:disperse}
One possible contributor to the population of sources outside the
prominent, obvious clusters in Figure~\ref{fig:galstrip} is the
dispersal of young stars from parent clusters or small groupings. The
infrared excess emission that allows us to identify candidate young
stars has a half life of around 2~Myr, although objects with an
infrared disk signature can still be seen in star forming regions as
old as 10~Myr \citep{aguilar06}. In 5~Myr, a Class II object traveling
at a typical velocity of 3~km~s$^{-1}$ \citep{furesz} would travel
about 15~pc. At a distance of 200~pc it would have moved about
4$\degr$ on the sky, at 1~kpc about 1$\degr$ and about 30$\arcmin$ at
a distance of 2~kpc. The combined lifetime for the Class I and Flat
SED phase of young stellar objects \citep{evans09} is roughly
1~Myr. At a typical velocity of 1~km~s$^{-1}$ \citep{offner09}, at a
distance of 200~pc it would have moved about 17$\arcmin$ on the sky,
or $\approx100\arcsec$ if at a distance of 2~kpc. Thus it is possible
that some of the most isolated young stars in the Outer Galaxy strip
are runaways from clusters, in particular the Class II sources. The
isolated Class I sources are harder to explain away completely in this
manner and some may be genuinely forming in very sparsely populated
environments. A more sophisticated model of the distribution of young
stars in the Outer Galaxy strip that incorporates cluster dispersal
would help us understand exactly how many stars may be forming in
isolation. Follow-up observations of the isolated Class I candidates
with higher spatial resolution and sensitivity and at other
wavelengths would also help understand how many of these objects are
cluster escapees or contaminant objects we have not previously
accounted for.

\section{Discussion}
\subsection{Filtering the {\it AllWISE} Catalog in the Galactic Plane}
The {\it AllWISE} photometric filtration scheme presented in this
paper is imperfect and ideally researchers would visually inspect {\it
  AllWISE} images of objects of interest found in the catalog in all
four bands. In the current era where big data is a powerful tool to
test theories in new and creative ways, this step may become
prohibitively time-consuming, thus the scheme presented here at least
allows a first pass to reduce the number of objects that need to be
manually checked. It is very important to note that different areas of
the Galactic Plane may behave differently than the star-forming test
regions analyzed here, particularly the Galactic Center region, where
confusion is likely to strongly affect performance. However, the
behavior of the various catalog parameters like signal to noise and
chi-squared in response to real and spurious sources and objects that
are confused or extended is generally applicable. It is certainly
important to note the relative strength and weakness of different
catalog parameters and we hope to have provided starting points for
researchers needing to find a pathway through the {\it AllWISE}
catalog to conduct the source searches and studies they care to do.

\subsection{Stellar Clustering}
Observations of the distributions of young stars in very large fields
with a survey like {\it WISE} enable us to take steps beyond previous
studies of star formation by allowing us to begin to connect the
observations made in nearby galaxies with those of individual star
forming regions in our own Milky Way. If a future, complete theory of
star formation is to be able to predict the star formation rate and
spatial distribution of stars in a large subsection of a galaxy given
the gas distribution and kinematics, then we need to have observations
of the large scale distribution of star formation in similar sized
regions against which to test such a theory. Our observations in this
paper, while limited by the properties of the {\it WISE} survey,
provide a beginning to the process of extracting and analyzing a full
census of star formation in a large portion of a spiral galaxy.

Our first observation of the distribution of the stars in our sample
extracted field shows an obvious degree of clustering into large
regions, but with a lower density population between them. We have
found that when considering this large scale distribution of young
stars in the Outer Galaxy, at most about 1/3 of stars can be fit with
a simple scale height distribution. We argue that this means that it
is a true statement that the majority of stars form in a clustered
environment, that is to say, in prominent massive star forming regions
whose density distribution is sharply peaked towards high
densities. However, we have also shown that, in a simple test to
correct for {\it WISE}'s incompleteness with respect to hypothetical
{\it Spitzer} Space Telescope observations, the size of the isolated
or non-clustered or scale height component may be a much smaller
fraction of the total distribution of stars, and may not be required
at all. The fact that the low density end of the distribution of stars
observed with {\it WISE} appears to be well fit with a scale height
similar to that of OB stars \citep[$\sim$50~pc,][]{reed00} and
molecular gas \citep[$\sim$70~pc,][]{cox05} in the Galactic Plane may
simply reflect the fact that {\it WISE} only detects a small fraction
of the true population of young stars. Since the {\it WISE} survey in
this paper is magnitude-limited, we tend to include distant, bright
objects and miss their possible neighboring cluster members. The
$\approx6\arcsec$ resolution of {\it WISE} also becomes significant
for more distant young stars. At 2~kpc 6$\arcsec$ equates to a
distance of 0.06~pc. At 3~kpc this resolution is 0.09~pc, which is
approaching the median separation of stars in Taurus \citep[about
  0.17~pc, from a simple computation of nearest neighbor distances for
  the sample of][]{hartmann05}. Thus the observed clustering
properties of more distant stars may reflect only the smooth, large
scale distribution of the gas and may miss their true nature as part
of the continuum of cluster sizes.

The largely hierarchical nature of the distribution of stars is also
reflected in the power law shape of the two-point correlation
function, $\xi$($\theta$). The break in the power law at roughly
2$\degr$ may also indicate a transition to isolated star formation. As
can be seen in our data, two degrees is approximately the size of
regions like W5, W3 and 4 and Cep~OB3. 

As described in $\S$~\ref{sec:disperse} it is possible that some
fraction of the apparently isolated young stars we see are the
dispersed members of young groups or clusters, in particular the Class
II sources and especially if there are a large quantity of objects
within 500~pc. The isolated protostars are harder to explain in this
way as dispersed cluster members, however since they are younger they
are more likely to be concentrated in physically small regions, like
the star-forming pillars around \htwo\ regions
\citep[e.g.][]{koenig08}. As we probe deeper, with higher resolution
observations and gain more information about stellar proper motions,
it may be that further observations of the low density population
reveal all young stars to be part of a single cluster continuum. We
intend to pursue this approach with a forthcoming study combining data
from the {\it Spitzer GLIMPSE360} Galactic Plane survey with {\it
  WISE} to examine the distribution of these stars when observed at
higher spatial resolution and sensitivity.

The work of \citet{bress10} and others has already shown that within
star forming regions, there is no obvious break in the density
distribution of young stars that would indicate a separate, isolated
component. If higher resolution and sensitivity observations do indeed
bear out the result that a truly distinct, isolated population is
either small or non-existent, this would be consistent with the work
of \citet{hopkins13} who argues that observed density distributions of
young stars are simply a continuum resulting from the imprint of the
hierarchical nature of supersonic turbulence in molecular clouds
(although as noted by Krumholz 2014, Klessen \& Burkert 2000 also
produce hierarchical structure in star forming gas without turbulence)
over all scales. In this picture, isolated (or at least, apparently
isolated) star formation would represent $< 10\%$ of the total in a
single, unified mechanism of star formation. If a truly isolated
population is supported by future observations, however, it will be
interesting to scrutinize its origins and mass function.

\section{Conclusions}
We have presented an assessment of the performance of {\it WISE} and
the {\it AllWISE} data release in a search for YSOs in a section of
the Galactic Plane. We have derived general results for the
completeness of the survey and this particular point source extraction
in the presence of modest clustering and bright nebular
backgrounds. We have also demonstrated the properties of real and
spurious detections in the {\it AllWISE} catalog and their behavior in
several key parameters provided in the catalog. We hope these will
provide guidance to researchers intending to use the point source
catalog in Galactic Plane regions.

We present one approach to improve the source reliability in the four
{\it WISE} bands with signal-to-noise and chi-squared cuts and we use
the resulting catalog to construct a new, revised young star detection
and classification scheme combining {\it WISE} and {\it 2MASS} near
and mid-infrared colors and magnitudes.

We apply this scheme to a section of the Outer Milky Way and assess
the clustering properties of the resulting distribution of Class I and
II stars using a 6th nearest neighbor density calculation and the
construction of the two-point correlation function. Our analysis
suggests that indeed, the majority of stars do form in massive star
forming regions, and any isolated mode of star formation is at most a
small fraction of the total star forming output of the Galaxy. We also
show that the isolated component may be very small and might thus
represent the tail end of a single mechanism of star formation in line
with models of molecular cloud collapse with supersonic turbulence and
not a separate mode all to itself.

In the near future, we will test these results with the addition of
higher resolution and sensitivity data from {\it GLIMPSE360} to
examine how the low density population is distributed and
clustered. We will also use available {\it Herschel} data to look at
the properties of the interstellar medium near these objects.

Further in the future still, measurements of the distances and
kinematics of these stars with the recently launched {\it Gaia} space
mission will allow us to overcome some of the limits of our magnitude
limited approach and give a more complete census of the diskless
population that may accompany the Class I and II sources we detect
here. Proper motion measurements made by {\it Gaia} of a significant
fraction of the young stellar objects seen in our survey will enable a
test of the fraction of the isolated population contributed by
dispersed and ejected cluster or group members.

\acknowledgements We thank an anonymous referee, whose comments and
suggestions improved the paper. Author Koenig gratefully acknowledges
support from NASA ADAP grant number NNX13AF07G. This work is based on
data obtained from (1) the {\it Wide-Field Infrared Survey Explorer},
which is a joint project of the University of California, Los Angeles,
and the Jet Propulsion Laboratory (JPL), California Institute of
Technology (Caltech), funded by the National Aeronautics and Space
Administration (NASA); (2) the Two Micron All Sky Survey,a joint
project of the University of Massachusetts and the Infrared Processing
and Analysis Center (IPAC)/Caltech, funded by NASA and the National
Science Foundation; and (3) the NASA/IPAC Infrared Science Archive,
which is operated by JPL, Caltech, under a contract with NASA. This
research has made use of NASA's Astrophysics Data System.

\appendix
\section{Appendix}
\subsection{Completeness}
\begin{figure}
\centering
\includegraphics[width=9cm]{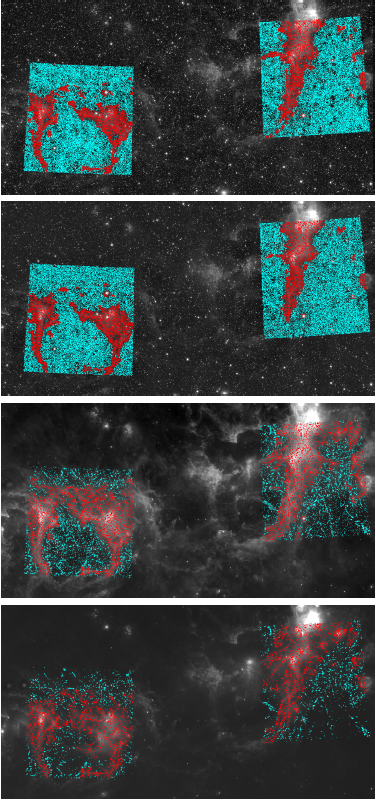}
\caption{Grayscale {\it WISE} images in the 4 bands overlaid with
  sources in the two test fields in those bands, colored coded for low
  sky (cyan points) and high sky (red points) following the thresholds
  in Table~\ref{tab:skycolr}.\label{fig:skycolr}}
\end{figure}

We derive the completeness, by band, on and off cloud (i.e. in regions
of high and low sky background) in two test regions covering W5 and
W3/W4. We define high and low sky regions by a threshold sky value (as
given in the {\it AllWISE} catalog) that as best as possible traces
the observed cloud boundaries in our {\it WISE} mosaic images when
catalog sources are over plotted and color-coded according to their
listed local sky value. These thresholds in sky background value in
digital numbers are listed in
Table~\ref{tab:skycolr}. Figure~\ref{fig:skycolr} shows plots of {\it
  AllWISE} catalog sources in the test fields color-coded in this way,
showing high (red) or low (blue) sky emission regions.

\begin{deluxetable}{ccc}
\tablewidth{0pt} 
\tabletypesize{\scriptsize}
\tablecaption{Sky Threshold Values}
\tablehead{\colhead{Band} & \colhead{W3} & \colhead{W5} }
\startdata
1 & 45 & 42 \\
2 & 55 & 50 \\
3 & 2050 & 2040 \\
4 & 555 & 562 
\label{tab:skycolr}
\enddata
\end{deluxetable}

\subsubsection{{\it WISE} Potential Completeness}
\begin{figure}
\centering
\includegraphics[width=17cm]{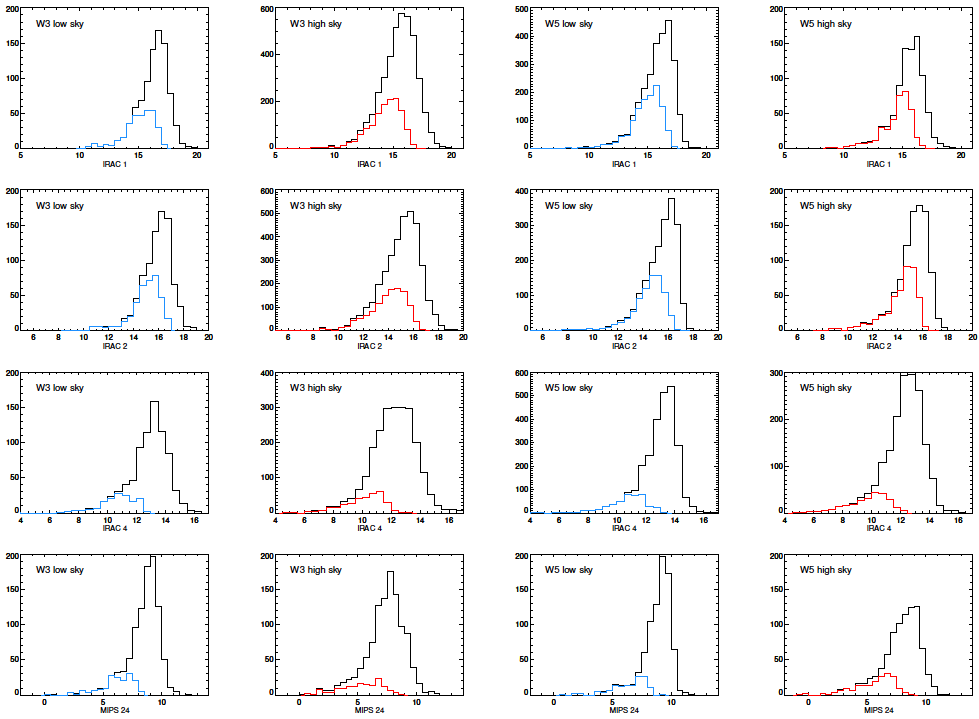}
\caption{Completeness histograms for the {\it WISE} truth catalog
  source lists against the {\it Spitzer} catalog (black histograms),
  in high and low sky regions (red and blue histograms respectively)
  in the two test regions.\label{fig:comphist}}
\end{figure}

\begin{figure}
\centering
\includegraphics[width=12cm]{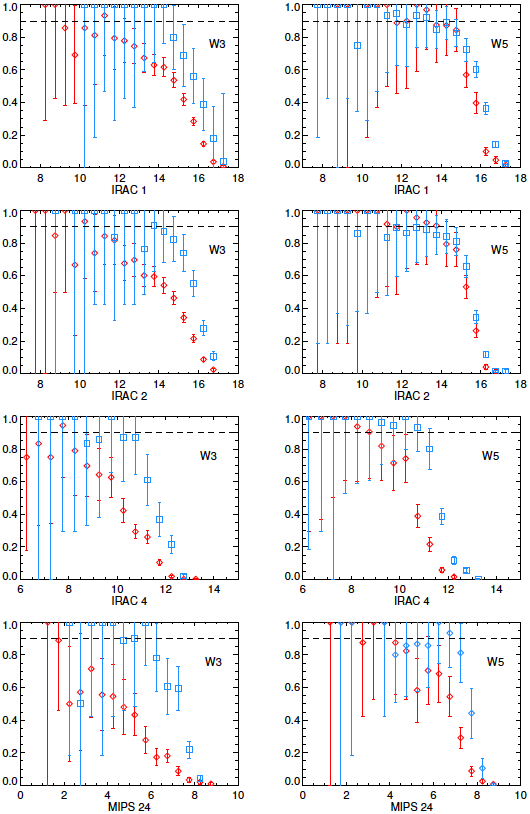}
\caption{Completeness plots derived from the histograms in
  Figure~\ref{fig:comphist} for the manual {\it WISE} truth catalog
  source lists.\label{fig:compplot} Red and blue points correspond to
  high sky and low sky regions respectively.}
\end{figure}

To derive a value for the completeness in each band, we extract {\it
  Spitzer} photometry from \citet{koenig08} in W5 and an extraction of
photometry from data in the {\it Spitzer} Heritage Archive for W3/W4
in IRAC bands 1, 2 and 4 and MIPS 24~$\micron$ corresponding roughly
to the four {\it WISE} bands respectively. Each source in the {\it
  Spitzer} lists is assigned the sky value of the nearest object in
the {\it AllWISE} catalog. We match our manually generated {\it WISE}
source list to the {\it Spitzer} catalogs with a 2$\arcsec$ search
radius to determine what fraction of the {\it Spitzer} objects could
be detected by {\it WISE}. Figure~\ref{fig:comphist} shows the
histograms of the {\it Spitzer} catalog and potential-{\it
  WISE}-retrieved subset, separated by field, by high or low sky
background and by band. We then plot the `detected' fraction (and
Poissonian error on the fraction) as a function of {\it Spitzer}
magnitude by field and band in Figure~\ref{fig:compplot}. The
magnitude where the error bar crosses the 90\% completeness line is
our completeness limit.

\subsubsection{{\it AllWISE} Completeness}
\begin{figure}
\centering
\includegraphics[width=17cm]{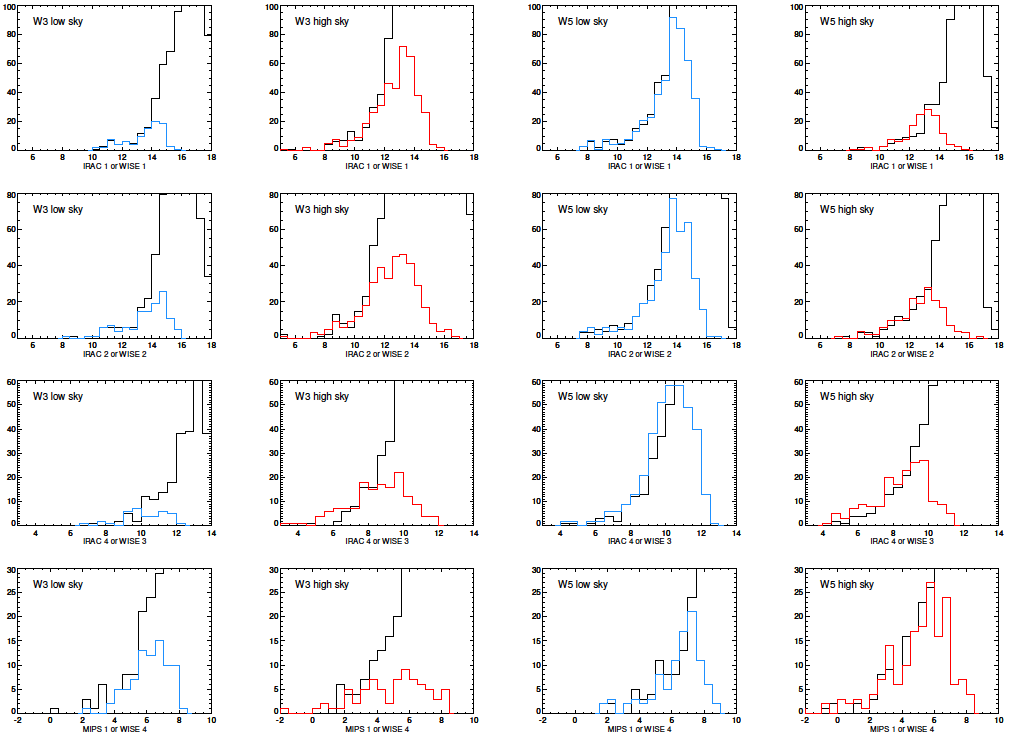}
\caption{Completeness histograms for {\it AllWISE} retrieved
  truth-catalog objects compared to the {\it Spitzer} catalog (black
  histograms) in high and low sky regions (red and blue histograms
  respectively) in the two test regions.\label{fig:allwcomphist}}
\end{figure}

\begin{figure}
\centering
\includegraphics[width=12cm]{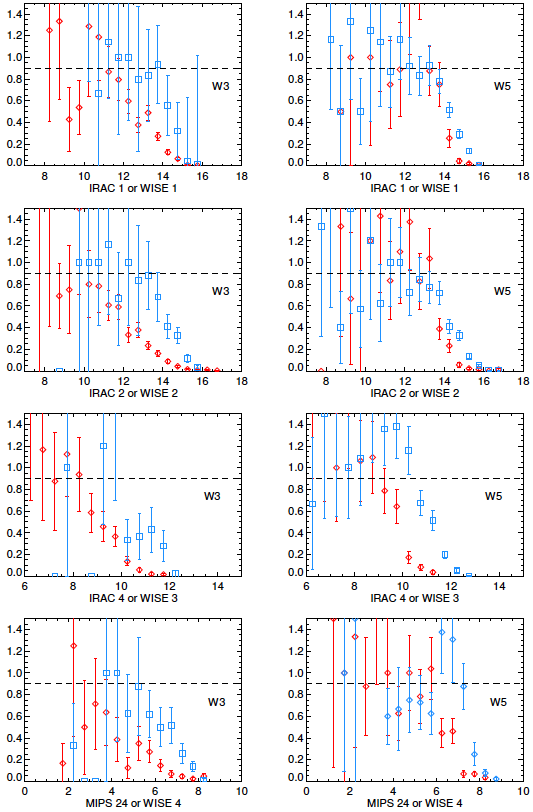}
\caption{Completeness plots derived from the histograms in
  Figure~\ref{fig:allwcomphist} for the {\it AllWISE} catalog source
  lists.\label{fig:awcompplot} Red and blue points correspond to high
  and low sky} regions respectively.
\end{figure}

The {\it AllWISE} source extraction pipeline is only able to retrieve
a subset of the sources visible in Galactic Plane {\it WISE}
images. We calculate the completeness of the {\it AllWISE} source
extraction by comparing histograms of source magnitude of {\it
  AllWISE} objects that match the truth catalog (so as to filter off
spurious detections), to the {\it Spitzer} catalogs for the same
regions. Figure~\ref{fig:allwcomphist} shows the histograms of the
{\it Spitzer} catalog and the {\it AllWISE} `real' sources over the
same area of the test regions. The mismatch in the histograms (for
example the W5 $w1$ high sky background plot in upper right) is
largely due to the binning in the plots and slight differences in the
magnitudes between {\it WISE} and {\it Spitzer}, but may also be due
to differences in how the background subtraction is handled or source
variability. In the $w3$ plots on the third row, {\it WISE} is able to
detect bright objects that are saturated in {\it Spitzer} images, the
bandpasses are less similar and spectral features \citep[e.g. the
  10~$\micron$ silicate feature, which can vary with evolutionary
  stage,][]{kessler} in these young objects' infrared spectra may
cause other mismatches. We plot the {\it AllWISE}-detected fraction
(and Poissonian error on the fraction) as a function of {\it Spitzer}
magnitude by field and band in Figure~\ref{fig:awcompplot}. The
magnitude where the error bar crosses the 90\% completeness line is
our completeness limit.

\subsection{Properties of Real and Fake Sources in the {\it AllWISE} Catalog}
Here we show the plots of the various catalog parameters analyzed in
this paper and discussed in $\S$~\ref{sec:overallmit} and their
behavior in the `real' by-eye source list and `fake' sources found in
the {\it AllWISE} Catalog. Black and red points show fake and real
sources respectively.

\begin{figure}
\centering
\includegraphics[width=13cm]{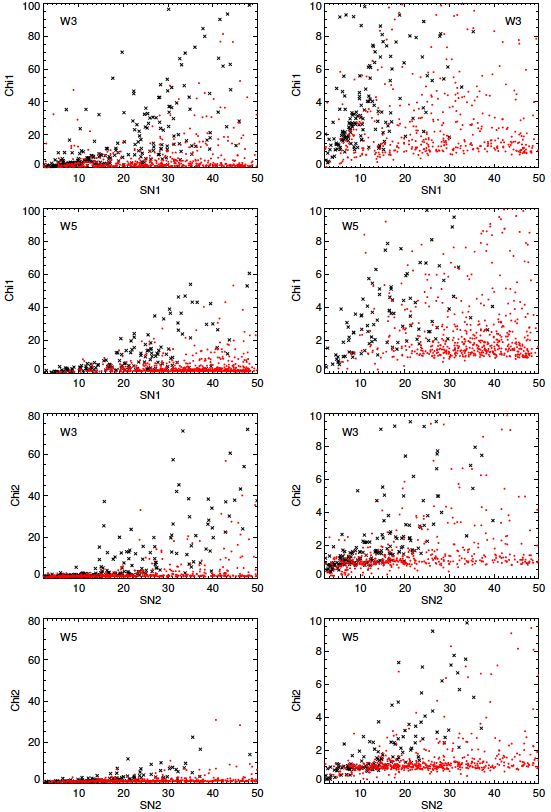}
\caption{Reduced chi-squared of the profile fit photometry versus
  signal to noise in sources that match an object in the {\it WISE}
  truth catalogs (red points) or have no visible match (black
  crosses).\label{fig:Appendix-SN-Chi}}
\end{figure}

Figure~\ref{fig:Appendix-SN-Chi} shows the behavior of $w?snr$ and
$w?rchi2$ in real and fake sources in the four WISE bands. The left
and right columns of plots show different zoom levels of the full
distribution in these parameters (left: zoomed out, right: zoomed
in). For each band we show the distribution separately between the two
test fields (as labeled in the upper corner of the plots). Real and
fake sources are found in overlapping, but different regions of this
parameter space that we use to separate the two in the {\it AllWISE}
catalog.

\begin{figure}
\centering
\includegraphics[width=13cm]{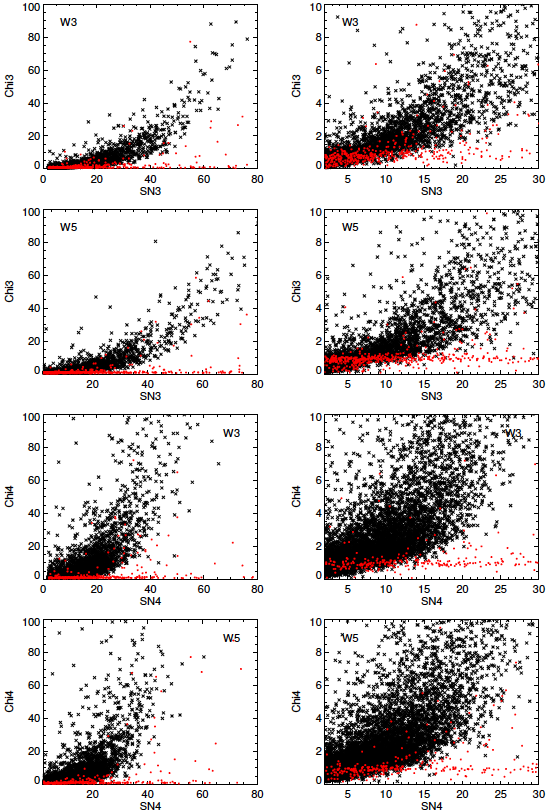}
\end{figure}

\begin{figure}
\centering
\includegraphics[width=12cm]{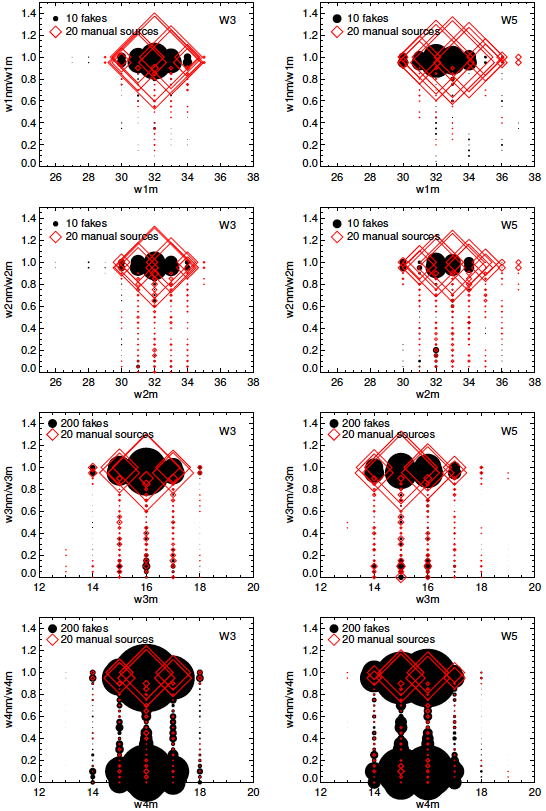}
\caption{Plots showing the ratio of N detections above $w?snr=3$ to N
  (all detections) versus N (all detections) for {\it AllWISE} sources
  in the two test regions. Red symbols correspond to matches to the
  truth catalog, black to {\it AllWISE} entries with no visible match
  in the {\it WISE} mosaic images. Symbol size scales with number of
  objects at that set of values.\label{fig:wnm-wm}}
\end{figure}

Figure~\ref{fig:wnm-wm} shows the behavior of $w?nm/w?m$ versus $w?m$
in `real' and `fake' sources in the four {\it WISE} bands, where
$w?nm$ is the number of profile-fit flux measurements for a source
with $w?snr > 3$ and $w?m$ is the number of profile-fit flux
measurements for a source. The symbol size scales with the number of
objects at that set of values for the two parameters as shown in the
figure legend. The left column of plots shows the behavior in the W3
test field, while the right column shows the W5 test field. For the
most part in bands 1, 2 and 3, both real and fake sources have similar
distributions of these parameters when plotted in this way. In band 4
a different pattern appears, with a much larger fraction of the fake
sources having $w4nm/w4m < 0.2$. The contrast between real and fake
source distributions is not strong enough to be used as an effective
discriminant in the catalog, however.

\begin{figure}
\centering
\includegraphics[width=13cm]{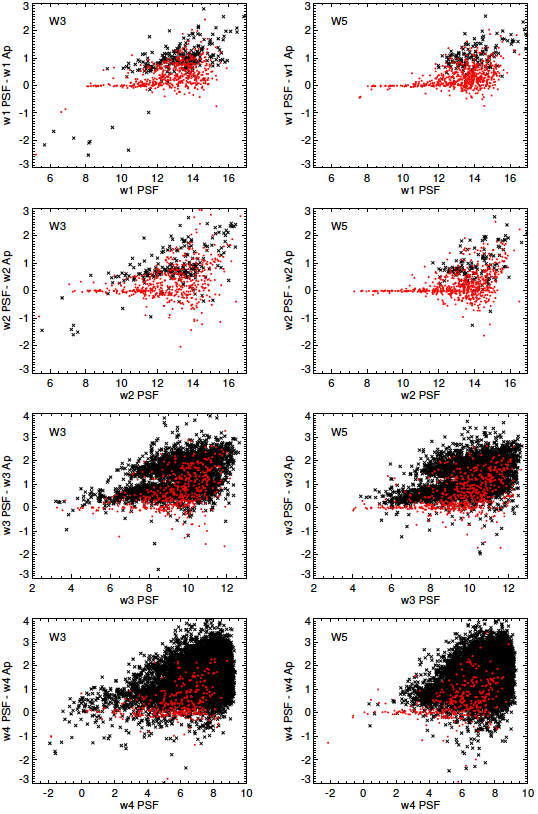}
\caption{Plots showing the difference between profile fit ($w?mpro$)
  and aperture ($w?mag$) magnitudes in all bands in {\it AllWISE}
  sources with a match in the truth catalogs (red points) and those
  with no match (black points).\label{fig:psf-ap}}
\end{figure}

Figure~\ref{fig:psf-ap} shows the behavior of the profile fit
($w?mpro$) and aperture ($w?mag$) magnitudes values for `real' and
`fake' sources in the {\it AllWISE} catalog in the two test fields. We
plot $w?mpro-w?mag$ versus $w?mpro$, with the left column of plots
showing the behavior in the W3 test field and the right column showing
the W5 test field. In bands 1 and 2, a cut of $|w?mpro-w?mag|<0.5$
would effectively remove the fake sources, while at the same time
allowing through the majority of the real objects (see
Table~\ref{tab:psf-ap}). A reader may validly pursue a cut based on
aperture and profile fit magnitude offsets in bands 1 and 2 as an
alternative to SNR vs. $\chi_{\nu}^2$. However, in bands 3 and 4, fake
sources are so numerous and occupy such a large portion of this
parameter space that no good cut to remove them and keep enough real
sources can be found to compete with our $w?snr$ and $w?rchi2$-based
cut in these bands.

\begin{thebibliography}{3}
\expandafter\ifx\csname natexlab\endcsname\relax\def\natexlab#1{#1}\fi

\bibitem[{{Acker} {et~al.}(1992){Acker}, {Marcout}, {Ochsenbein}, {Stenholm}, \& {Tylenda}}]{acker92}
{Acker}, A., {Marcout}, J., {Ochsenbein}, F., {Stenholm}, B., \& {Tylenda}, R. 1992, Strasbourg-ESO Catalogue of Galactic Planetary Nebulae, Parts I, II (Garching: European Southern Observatory)

\bibitem[{{Alksnis} {et~al.}(2001){Alksnis}, {Balklavs}, {Dzervitis}, {Eglitis}, {Paupers}, \& {Pundure}}]{alks}
{Alksnis}, A., {Balklavs}, A., {Dzervitis}, U., et al.\ 2001, VizieR Online Data Catalog, 3227, 0

\bibitem[{{Allen} {et~al.}(2004){Allen}, {Calvet}, {D'Alessio}, {Merin}, {Hartmann}, {Megeath}, {Gutermuth}, {Muzerolle}, {Pipher}, {Myers}, \& {Fazio}}]{allen04}
{Allen}, L.~E., {Calvet}, N., {D'Alessio}, P. et~al.\ 2004, \apjs, 154, 363

\bibitem[{{Allen} {et~al.}(2012){Allen}, {Gutermuth}, {Kryukova}, {Megeath}, {Pipher}, {Naylor}, {Jeffries}, {Wolk}, {Spitzbart}, \& {Muzerolle}}]{allen12}
{Allen}, T.~S., {Gutermuth}, R.~A., {Kryukova}, E., et~al.\  2012, \apj, 750, 125

\bibitem[Andr{\'e} et al.(2010)]{andre10}
Andr{\'e}, P., Men'shchikov, A., Bontemps, S., et al.\ 2010, \aap, 518, L102

\bibitem[{{Andrews} {et~al.}(2011){Andrews}, {Wilner}, {Espaillat}, {Hughes}, {Dullemond}, {McClure}, {Qi}, \& {Brown}}]{andrews11}
{Andrews}, S.~M., {Wilner}, D.~J., {Espaillat}, C., et~al.\ 2011, \apj, 732, 42

\bibitem[{{Assef} {et~al.}(2010){Assef}, {Kochanek}, {Brodwin}, {Cool}, {Forman}, {Gonzalez}, {Hickox}, {Jones}, {Le Floc'h}, {Moustakas}, {Murray}, \& {Stern}}]{assef10}
{Assef}, R.~J., {Kochanek}, C.~S., {Brodwin}, M., et al.\ 2010, \apj, 713, 970

\bibitem[{{Bressert} {et~al.}(2010){Bressert}, {Bastian}, {Gutermuth}, {Megeath}, {Allen}, {Evans}, {Rebull}, {Hatchell}, {Johnstone}, {Bourke}, {Cieza}, {Harvey}, {Merin}, {Ray}, \& {Tothill}}]{bress10}
{Bressert}, E., {Bastian}, N., {Gutermuth}, R. 2010, \mnras, 409, L54

\bibitem[{{Carey} {et~al.}(1998){Carey}, {Clark}, {Egan}, {Price}, {Shipman}, \& {Kuchar}}]{carey98}
{Carey}, S.~J., {Clark}, F.~O., {Egan}, M.~P., et~al.\ 1998, \apj, 508, 721

\bibitem[{{Carpenter}(2000)}]{carpenter00}
{Carpenter}, J.~M.\ 2000, \aj, 120, 3139

\bibitem[{{Casertano} \& {Hut}(1985)}]{casertano85}
{Casertano}, S., \& {Hut}, P. 1985, \apj, 298, 80

\bibitem[{{Chengalur} {et~al.}(1996){Chengalur}, {Lewis}, {Eder}, \& {Terzian}}]{chenga}
{Chengalur}, J.~N., {Lewis}, B.~M., {Eder}, J., \& {Terzian}, Y. 1996, VizieR Online Data Catalog, 208, 1

\bibitem[{{Cieza} {et~al.}(2012){Cieza}, {Schreiber}, {Romero}, {Williams}, {Rebassa-Mansergas}, \& {Mer{\'{\i}}n}}]{cieza12}
{Cieza}, L.~A., {Schreiber}, M.~R., {Romero}, G.~A., et~al.\ 2012, \apj, 750, 157

\bibitem[{{Cohen}(1993)}]{cohen93}
{Cohen}, M.\ 1993, \aj, 105, 1860

\bibitem[{{Cohen} {et~al.}(2011){Cohen}, {Parker}, {Green}, {Miszalski}, {Frew}, \& {Murphy}}]{cohen11}
{Cohen}, M., {Parker}, Q.~A., {Green}, A.~J., et~al.\ 2011, \mnras, 413, 514

\bibitem[{{Cox}(2005)}]{cox05}
{Cox}, D.~P.\ 2005, \araa, 43, 337

\bibitem[{{de Wit} {et~al.}(2005){de Wit}, {Testi}, {Palla}, \& {Zinnecker}}]{dewit05}
{de Wit}, W.~J., {Testi}, L., {Palla}, F., \& {Zinnecker}, H. 2005, \aap, 437, 247

\bibitem[{{Dominik} \& {Decin}(2003)}]{dominik}
{Dominik}, C., \& {Decin}, G.\ 2003, \apj, 598, 626

\bibitem[{{Evans} {et~al.}(2009){Evans}, {Dunham}, {J{\o}rgensen}, {Enoch}, {Mer{\'{\i}}n}, {van Dishoeck}, {Alcal{\'a}}, {Myers}, {Stapelfeldt}, {Huard}, {Allen}, {Harvey}, {van Kempen}, {Blake}, {Koerner}, {Mundy}, {Padgett}, \& {Sargent}}]{evans09}
{Evans}, N.~J., {Dunham}, M.~M., {J{\o}rgensen}, J.~K., et~al.\ 2009, \apjs, 181, 321

\bibitem[{{F{\H u}r{\'e}sz} {et~al.}(2006){F{\H u}r{\'e}sz}, {Hartmann}, {Szentgyorgyi}, {Ridge}, {Rebull}, {Stauffer}, {Latham}, {Conroy}, {Fabricant}, \& {Roll}}]{furesz}
{F{\H u}r{\'e}sz}, G., {Hartmann}, L.~W., {Szentgyorgyi}, A.~H., et~al. 2006, \apj, 648, 1090

\bibitem[{{Greene} {et~al.}(1994){Greene}, {Wilking}, {Andr{\'e}}, {Young}, \& {Lada}}]{greene94}
{Greene}, T.~P., {Wilking}, B.~A., {Andr{\'e}}, P., {Young}, E.~T., \& {Lada}, C.~J. 1994, \apj, 434, 614

\bibitem[{{Gutermuth} {et~al.}(2009){Gutermuth}, {Myers}, {Megeath}, {Allen}, {Pipher}, {Muzerolle}, {Porras}, {Winston}, \& {Fazio}}]{gutermu08}
{Gutermuth}, R.~A., {Myers}, P.~C., {Megeath}, S.~T., et~al.\ 2008, \apj, 674, 336

\bibitem[{{Gutermuth} {et~al.}(2009){Gutermuth}, {Megeath}, {Myers}, {Allen}, {Pipher}, \& {Fazio}}]{gutermu09}
{Gutermuth}, R.~A., {Megeath}, S.~T., {Myers}, P.~C., et~al.\ 2009, \apjs, 184, 18

\bibitem[Hartmann et al.(2005)]{hartmann05} 
Hartmann, L.~W., Megeath, S.~T., Allen, L., et al.\ 2005, \apj, 629, 881

\bibitem[{{Hillwig} {et~al.}(2006){Hillwig}, {Gies}, {Bagnuolo}, {Huang}, {McSwain}, \& {Wingert}}]{hillwig}
{Hillwig}, T.~C., {Gies}, D.~R., {Bagnuolo}, Jr., W.~G., et~al.\ 2006, \apj, 639, 1069

\bibitem[{{Hopkins}(2013)}]{hopkins13}
{Hopkins}, P.~F.\ 2013, \mnras, 428, 1950

\bibitem[{{Kessler-Silacci} {et~al.}(2005){Kessler-Silacci} {Hillenbrand}, {Blake}, \& {Meyer}}]{kessler}
{Kessler-Silacci}, J.~E., {Hillenbrand}, L.~A., {Blake}, G.~A., \& {Meyer}, M.~R. 2005, \apj, 622, 404

\bibitem[{{Klessen} \& {Burkert}(2000)}]{klessen}
{Klessen}, R.~S., \& {Burkert}, A. 2000, \apjs, 128, 287

\bibitem[{{Koenig} {et~al.}(2008){Koenig}, {Allen}, {Gutermuth}, {Hora}, {Brunt}, \& {Muzerolle}}]{koenig08}
{Koenig}, X.~P., {Allen}, L.~E., {Gutermuth}, R.~A., et~al.\ 2008, \apj, 688, 1142

\bibitem[{{Koenig} {et~al.}(2012){Koenig}, {Leisawitz}, {Benford}, {Rebull}, {Padgett}, \& {Assef}}]{koenig12}
{Koenig}, X.~P., {Leisawitz}, D.~T., {Benford} D.~J., et~al.\ 2012, \apj, 744, 130

\bibitem[{{Kraus} \& {Hillenbrand}(2008)}]{kraus}
{Kraus}, A.~L., \& {Hillenbrand}, L.~A.\ 2008, \apjl, 686, L111

\bibitem[{{Krumholz}(2014)}]{krumholz}
{Krumholz}, M.~R.\ 2014, \physrep, in press, arXiv1402.0867

\bibitem[{{Larson}(1995)}]{larson}
{Larson}, R.~B.\ 1995, \mnras, 272, 213

\bibitem[{{Li} \& {Draine}(2001)}]{li01}
{Li}, A., \& {Draine}, B.~T.\ 2001, \apj, 554, 778

\bibitem[{{Lindqvist} {et~al.}(1992){Lindqvist}, {Winnberg}, {Habing}, \& {Matthews}}]{lindq}
{Lindqvist}, M., {Winnberg}, A., {Habing}, H.~J., \& {Matthews}, H.~E. 1992, \aaps, 92, 43

\bibitem[{{McClure}(2009)}]{mcclure09}
{McClure}, M.\ 2009, \apj, 693, 81

\bibitem[{{Marsh} \& {Jarrett}(2012)}]{marsh12}
{Marsh}, K.~A., \& {Jarrett}, T.~H.\ 2012, \pasa, 29, 269

\bibitem[{{Masiunas} {et~al.}(2012){Masiunas}, {Gutermuth}, {Pipher}, {Megeath}, {Myers}, {Allen}, {Kirk}, \& {Fazio}}]{masiunas12}
{Masiunas}, L.~C., {Gutermuth}, R.~A., {Pipher}, J.~L., et~al.\ 2012, \apj, 752, 127

\bibitem[{{Mathew} {et~al.}(2008){Mathew}, {Subramaniam}, \& {Bhatt}}]{mathew}
{Mathew}, B., {Subramaniam}, A., \& {Bhatt}, B.~C. 2008, \mnras, 388, 1879

\bibitem[{{Miszalski} {et~al.}(2008){Miszalski}, {Parker}, {Acker}, {Birkby}, {Frew}, \& {Kovacevic}}]{miszalski08}
{Miszalski}, B., {Parker}, Q.~A., {Acker}, A., et~al.\ 2008, \mnras, 384, 525

\bibitem[Morales et al.(2012)]{morales} 
Morales, F.~Y., Padgett, D.~L., Bryden, G., Werner, M.~W., \& Furlan, E.\ 2012, \apj, 757, 7 

\bibitem[Nikutta et al.(2014)]{nikutta} 
Nikutta, R., Hunt-Walker, N., Nenkova, M., Ivezi{\'c}, {\v Z}., \& Elitzur, M.\ 2014, arXiv:1405.7966 

\bibitem[Offner et~al.(2009){Offner}, {Hansen}, \& {Krumholz}]{offner09} 
Offner, S.~S.~R., Hansen, C.~E., \& Krumholz, M.~R.\ 2009, \apjl, 704, L124

\bibitem[{{Parker} {et~al.}(2006){Parker}, {Acker}, {Frew}, {Hartley}, {Peyaud}, {Ochsenbein}, {Phillipps}, {Russeil}, {Beaulieu}, {Cohen}, {K{\"o}ppen}, {Miszalski}, {Morgan}, {Morris}, {Pierce}, \& {Vaughan}}]{parker06}
{Parker}, Q.~A., {Acker}, A., {Frew}, D.~J., et~al.\ 2006, \mnras, 373, 79

\bibitem[{{Peebles} \& {Hauser}(1974)}]{peebles74}
{Peebles}, P.~J.~E., \& {Hauser}, M.~G.\ 1974, \apjs, 28, 19

\bibitem[{{P{\'e}rault} {et~al.}(1996){P{\'e}rault}, {Omont}, {Simon}, {Seguin}, {Ojha}, {Blommaert}, {Felli}, {Gilmore}, {Guglielmo}, {Habing}, {Price}, {Robin}, {de Batz}, {Cesarsky}, {Elbaz}, {Epchtein}, {Fouque}, {Guest}, {Levine}, {Pollock}, {Prusti}, {Siebenmorgen}, {Testi}, \& {Tiphene}}]{perault}
{P{\'e}rault}, M., {Omont}, A., {Simon}, G., et~al.\ 1996, \aap, 315, L165

\bibitem[{{Polychroni} {et~al.}(2010){Polychroni}, {Moore}, \& {Allsopp}}]{polychron10}
{Polychroni}, D., {Moore}, T.~J.~T., \& {Allsopp}, J. 2010, in ASP
Conf. Ser. 424, 9th International Conference of the Hellenic
Astronomical Society, ed. K.  Tsinganos, D. Hatzidimitriou, \&
T. Matsakos (San Francisco, CA: ASP), 165

\bibitem[{{Rathborne} {et~al.}(2006)}]{rathb06}
{Rathborne}, J.~M., {Jackson}, J.~M., \& {Simon}, R. 2006, \apj, 641, 389

\bibitem[{{Rebull} {et~al.}(2010){Rebull}, {Padgett}, {McCabe}, {Hillenbrand}, {Stapelfeldt}, {Noriega-Crespo}, {Carey}, {Brooke}, {Huard}, {Terebey}, {Audard}, {Monin}, {Fukagawa}, {G{\"u}del}, {Knapp}, {Menard}, {Allen}, {Angione}, {Baldovin-Saavedra}, {Bouvier}, {Briggs}, {Dougados}, {Evans}, {Flagey}, {Guieu}, {Grosso}, {Glauser}, {Harvey}, {Hines}, {Latter}, {Skinner}, {Strom}, {Tromp}, \& {Wolf}}]{rebull10}
{Rebull}, L.~M., {Padgett}, D.~L., {McCabe}, C.-E., et~al.\ 2010, \apjs, 186, 259

\bibitem[{{Reed}(2000)}]{reed00}
{Reed}, B.~C.\ 2000, \aj, 120, 314

\bibitem[{{Rivinius} {et~al.}(2013){Rivinius}, {Carciofi}, \& {Martayan}}]{rivin13}
{Rivinius}, T., {Carciofi}, A.~C., \& {Martayan}, C. 2013, \aapr, 21, 69

\bibitem[{{Roberts}(1957)}]{roberts57}
{Roberts}, M.~S. 1957, \pasp, 69, 59

\bibitem[{Robitaille {et~al.}(2008){Robitaille}, {Meade}, {Babler}, {Whitney}, {Johnston}, {Indebetouw}, {Cohen}, {Povich}, {Sewilo}, {Benjamin}, \& {Churchwell}}]{robitaille08} 
Robitaille, T.~P., Meade, M.~R., Babler, B.~L., et al.\ 2008, \aj, 136, 2413 

\bibitem[{{Schlegel} {et~al.}(1998)}]{schlegel98}
{Schlegel}, D.~J., {Finkbeiner}, D.~P., \& {Davis}, M. 1998, \apj, 500, 525

\bibitem[{{Sicilia-Aguilar} {et~al.}(2006){Sicilia-Aguilar}, {Hartmann}, {Calvet}, {Megeath}, {Muzerolle}, {Allen}, {D'Alessio}, {Mer{\'{\i}}n}, {Stauffer}, {Young}, \& {Lada}}]{aguilar06}
{Sicilia-Aguilar}, A., {Hartmann}, L.~W., {Calvet}, N., et~al.\ 2006, \apj, 638, 897

\bibitem[{{Simon}(1997)}]{simon97}
{Simon}, M.\ 1997, \apjl, 482, 81

\bibitem[{{Sjouwerman} {et~al.}(1998){Sjouwerman}, {van Langevelde}, {Winnberg}, \& {Habing}}]{sjouw}
{Sjouwerman}, L.~O., {van Langevelde}, H.~J., {Winnberg}, A., \& {Habing}, H.~J. 1998, \aaps, 128, 35

\bibitem[{{Skrutskie} {et~al.}(2006){Skrutskie}, {Cutri}, {Stiening},
  {Weinberg}, {Schneider}, {Carpenter}, {Beichman}, {Capps}, {Chester},
  {Elias}, {Huchra}, {Liebert}, {Lonsdale}, {Monet}, {Price}, {Seitzer},
  {Jarrett}, {Kirkpatrick}, {Gizis}, {Howard}, {Evans}, {Fowler}, {Fullmer},
  {Hurt}, {Light}, {Kopan}, {Marsh}, {McCallon}, {Tam}, {Van Dyk}, \&
  {Wheelock}}]{skrut06}
{Skrutskie}, M.~F., {Cutri}, R.~M., {Stiening}, R., et~al.\ 2006, \aj, 131, 1163

\bibitem[{{Wainscoat} {et~al.}(1992){Wainscoat}, {Cohen}, {Volk}, {Walker}, \& {Schwartz}}]{wainscoat92}
{Wainscoat}, R.~J., {Cohen}, M., {Volk}, K., {Walker}, H.~J., \& {Schwartz}, D.~E. 1992, \apjs, 83, 111

\bibitem[{{Westerhout}(1958)}]{westerhout58}
{Westerhout}, G.\ 1958, \bain, 14, 215

\bibitem[{{Wright} {et~al.}(2010){Wright}, {Eisenhardt}, {Mainzer}, {Ressler}, {Cutri}, {Jarrett}, {Kirkpatrick}, {Padgett}, {McMillan}, {Skrutskie}, {Stanford}, {Cohen}, {Walker}, {Mather}, {Leisawitz}, {Gautier}, {McLean}, {Benford}, {Lonsdale}, {Blain}, {Mendez}, {Irace}, {Duval}, {Liu}, {Royer}, {Heinrichsen}, {Howard}, {Shannon}, {Kendall}, {Walsh}, {Larsen}, {Cardon}, {Schick}, {Schwalm}, {Abid}, {Fabinsky}, {Naes}, \& {Tsai}}]{wright10}
{Wright}, E.~L., {Eisenhardt}, P.~R.~M., {Mainzer}, A.~K., et~al.\ 2010, \aj, 140, 1868

\bibitem[{{Zinnecker} \& {Yorke}(2007)}]{zinnecker07}
{Zinnecker}, H., \& {Yorke}, H.~W.\ 2007, \araa, 45, 481

\end{thebibliography}
\end{document}